\documentclass[]{ptephy_v1}
\usepackage{amsmath}
\usepackage{color}
\usepackage{url}
\usepackage{graphicx}
\preprintnumber{XXXX-XXXX} 
\newcommand{\h}{\mathchar`-}

\newcommand{\prl}{Phys. Rev. Lett.}
\newcommand{\apj}{Astrophys. J.}
\newcommand{\apjl}{Astrophys. J. Lett.}
\newcommand{\nat}{Nature}
\newcommand{\araa}{Annu. Rev. }
\newcommand{\prd}{Phys. Rev. D}
\newcommand{\aap}{Astron. Astrophys.}
\newcommand{\mnras}{Mon. Not. R. Astron. Soc.}
\newcommand{\baas}{BAAS}
\newcommand{\ssr}{Space sci. Rev.SR}
\newcommand{\aj}{Astron. J.}
\newcommand{\apjs}{Astrophys. J. Suppl.}
\newcommand{\pasj}{Publ. Asron. Soc. Jpn.}
\newcommand{\aplett}{Astrophys. Lett.}
\begin{document}

\title{Analytic properties of electromagnetic field of binary compact stars and electromagnetic precursors to gravitational waves}

\author[1,2,*]{Tomoki Wada}
\author[1,3]{Masaru Shibata}
\author{Kunihito Ioka}
\affil[1]{Center for Gravitational Physics,  Yukawa Institute for Theoretical Physics, Kyoto University, Kyoto, 606-8502, Japan\email{tomoki.wada@yukawa.kyoto-u.ac.jp}}
\affil[2]{JSPS Research Fellow}
\affil[3]{Max Planck Institute for Gravitational Physics (Albert Einstein Institute), Am M\"{u}hlenberg 1, Potsdam-Golm, 14476, Germany}

\begin{abstract}
We analytically study the properties of the electromagnetic field in vacuum around close binary compact stars containing at least one neutron star.
 We show that the orbital motion of the neutron star induces high multipole modes of the electromagnetic field just before the merger.
 These modes are superimposed to form a spiral arm configuration and its edge is found to be a likely site for magnetic reconnection.
 These modes also enhance the total Poynting flux from neutron star binaries by a factor of 2--4.
 We also indicate that the electric field induced by the orbital motion lead to a magnetosphere around binaries and estimate its plasma density, which has a different parameter dependence than Goldreich-Julian density.
 With these properties, we discuss possible electromagnetic counterparts to gravitational wave events, and identify radio precursors, such as fast radio bursts, as the most promising observational targets.
\end{abstract}
\subjectindex{E01, E02, E32, E38}
\maketitle

\section{Introduction}
%back ground
In August 2017, gravitational waves from a binary neutron star (BNS) merger, GW170817, and its electromagnetic counterparts, GRB170817A \& AT2017 gfo, were observed by aLIGO/VIRGO and many telescopes \citep{LIGO2017gw,LIGO2017multi,LIGO2017gwgrb}.
 This event had a lot of implications on astrophysics and on gravitational physics; for example, the origin of short gamma-ray bursts \citep{Pac1986,Goo1986,EicLiv1989}, the origin of \(r\)-process elements \citep{LatSch1974}, the estimation of Hubble constant \citep{Sch1986,HotNak2019}, and the speed of gravitational waves \citep{LIGO2018ns,LIGO2017Hubble}.
 This discovery showed that multi-messenger astronomy of gravitational-wave and electromagnetic-wave observations contributes to answering many of the unsolved issues.
 In the aLIGO/VIRGO O3 run, one gravitational wave event, GW190425, was reported so far \citep{LIGO2020GW190425}.
 There are several other candidates for BNS mergers and black hole - neutron star (BH-NS) mergers.
 
%mention about EMCP 
Many models have been suggested for electromagnetic counterparts of neutron-star (NS) mergers (for review, \citep{BarBra2013,Ros2015,Fer2016}).
 A short gamma-ray burst \citep{Pac1986,Ber2014}, its afterglow \citep{SarPir1998}, and a kilonova/macronova \citep{LiPac1998,Met2019} have been detected in GW170817.
 However, there might be other electromagnetic counterparts.
 One candidate is a late time engine activity which is driven by the fall-back accretion to the central black hole (BH) \citep{KisIok2015} or by the spin-down luminosity of the magnetar formed after the merger of a BNS.
 Another candidate is an electromagnetic precursor, which is driven by the orbital motion or by the magnetospheric interaction of NSs. 

%precursor
There are several precursor models, dipole radiation \citep{IokTan2000}, dipole radiation from charged compact binaries \citep{Zha2019}, DC circuit (unipolar induction) \citep{McwLev2011,Lyu2011,Lyu2011II,Lai2012,Pir2012,DorLev2016}, interaction of magnetospheres \citep{Vie1996,HanLyu2001,PalLeh2013,PalLeh2013sup,WanPen2018,Lyu2018,CriCer2019,MosPhi2020}, and fireball model \citep{MetZiv2016,DorLev2016}.
 The physical process that takes place before the merger and the physical quantities, such as  the strength of the magnetic field of the NS(s) in binaries, are not well understood.
 Furthermore, because the electromagnetic precursor is radiated before the merger,  it is not easy to detect it.
 However, in the era of the space-based gravitational wave detector, Laser Interferometer Space Antenna (LISA) \citep{LISA2017}, the sky position will be predicted before the merger \citep{TakNak2003} and then the electromagnetic precursor will be a new target for the observations.
 If DECi-hertz Interferometer Gravitational wave Observatory (DECIGO) \citep{DECIGO2006,DECIGO2017}, or similar detectors that have sensitivity below $10\,{\rm Hz}$, such as TaiJi \citep{TaiJi2018} and TianQin \citep{TianQin2016}, are realized, we will have more opportunities to observe electromagnetic precursors of NS mergers.

% show topics
In this paper, we analytically study the properties of the electromagnetic field in the vacuum around a binary system of compact stars containing at least one NS. 
 In compact NS binaries, the velocity of the compact objects just before the merger is 30--40\% of the speed of light, and we have to take the special relativistic effects on the electromagnetic field into account.
 Motivated by this fact, we analytically solve Maxwell equations in vacuum using vector harmonics expansion, which retains the special relativistic correction in a straightforward manner.
 First, we show the analytic solution for the vacuum electromagnetic field induced by a moving magnetic dipole, and display the configuration of the electromagnetic field.
 With this analytic solution, we calculate the total Poynting flux radiated from binaries and show that the special relativistic effect enhances the Poynting flux.
 Second, we indicate that the orbital motion of a NS can induce a magnetosphere around the binary, and estimate the number density of the plasma to explore its plasma effect on the electromagnetic precursor of binary mergers.
 Third, we discuss the observability of the precursor of binary mergers assuming that the emission would be similar to that from pulsars.

%sabetuka
In this study, we only consider the vacuum case without plasma.
 In some numerical simulations, the plasma effects have been taken into account.
 For example,  a resistive magnetohydrodynamic simulation for the electromagnetic field around binary compact stars have been performed in Palenzuela et al. (2013)\citep{PalLeh2013sup} and Palenzuela et al. (2013)\citep{PalLeh2013}.
 A particle-in-cell simulation without the orbital motion of binaries have been performed in Crinquand et al. (2019) \cite{CriCer2019}.
 Force-free simulations of a NS binary have been performed in Carrasco and Shibata (2020) \citep{CarShi2020} and Most and Philippov (2020) \citep{MosPhi2020}.
 These simulations show the luminosity from binaries with plasma for a few milliseconds before the merger.
 We analytically derive a radiation formula which clarifies the dependence of the luminosity on the orbital separation of binaries and on other parameters (e.g., mass, radius, and magnetic field of the compact stars).
 Thus, our work is complementary to these previous works.
 Also, our calculation contains the special relativistic effect that enhances the Poynting flux from the binary especially just before the merger.
 In Ioka and Taniguchi (2000) \citep{IokTan2000}, the Poynting flux is calculated in zeroth order of $v/c$ (where $v$ is the velocity of the NS and $c$ is the speed of light).
 Our analytic treatment makes it possible to calculate the Poynting flux in the series of $v/c$.
 These higher order terms enhance the Poynting flux by a factor of $2$--$4$.
 Our analytic results are consistent with the numerical result of Carrasco and Shibata (2020) \citep{CarShi2020}.
 
%construction
The paper is organized as follows.
 In section \ref{sec:2}, we show the exact solution of the electromagnetic field induced by a magnetic dipole moment orbiting around the companion (NS or BH).
 Motivated by the fact that some pulsars are considered to release particle wind, or to radiate coherent radio waves via magnetic reconnection process in their current sheets \citep{CheBel2014,Lyb2019,PhiUzd2019}, we speculate the location of magnetic reconnection induced by the orbital motion of the NS if plasma exists.
 We also evaluate the effect of the higher multipole modes on the total luminosity.
 In section \ref{sec:3}, we first indicate that the orbital motion of a magnetized NS induces a magnetosphere around the NS.
 Then, we study the observability of the precursor as an electromagnetic counterpart of binary mergers.
 Section \ref{sec:4} is devoted for summary and discussion.

%%%
%%%section
%%%
\section{Electromagnetic field around a NS binary}
\label{sec:2}
%overview of magnetosphere creation
In this section, we consider the electromagnetic field induced by the orbital motion of a magnetized NS in compact binary systems.
 We consider the case in which the binary contains at least one magnetized NS and is in a late inspiral phase about a few milliseconds before the merger.
 For simplicity, we assume that the NS is orbiting around its companion in Kepler motion.
 We also assume that the NS does not have its intrinsic spin.
 The former assumption is reasonable during the inspiral phase, that is, until the binary separation reaches the innermost stable circular orbit (ISCO) of the companion \citep{Maggiore2008,Shibata2016N}.
 The latter is justified in the late inspiral phase.
 This is because tidal locking is not effective and, just before the merger, the orbital velocity is usually much higher that the spin velocity of the NS \citep{BilCut1992}.
 Also, we use a dipole approximation for the magnetic field of the NS.
 In this section, we use units in which $G=1$ and $c=1$ where $G$ is the gravitational constant.

%position of NS
With these assumptions, we consider the electromagnetic field induced by the orbiting magnetic dipole moment.
 The mass of the NS is \(M_{\rm{NS}}\), the mass of the companion is \(M_{\rm{c}}\), and the separation of the binary is $R$.
 Also, the total mass is $M(= M_{\rm NS}+M_c)$ and the distance between the NS and the center-of-mass of the binary is $a(=RM_c/M)$.
 The binary orbital motion is in the \(x\h y\) plane. 
 Then, in Cartesian coordinates, the NS is at
\begin{equation}
\vec{r}_{\rm{NS}}(t)=(a\cos{\Omega t},\, a\sin{\Omega t},\,0),
\end{equation}
where $\Omega=\sqrt{\frac{M_{\rm NS}+M_{\rm c}}{R^3}}$ is the angular velocity and \(t\) is the time in the center-of-mass frame.

%%%
%subsection
\subsection{Vector harmonics expansion and the exact solution of electromagnetic field}
%Maxwell equation
The definition of vector spherical harmonics expansion of a vector \(\vec{B}\) is \citep{RufTio1972}
\begin{eqnarray}
\vec{B}=\sum_{l,m} \left[B_1^{lm} \biggl(Y_{lm}(\theta,\phi),0,0\biggr)+B_2^{lm}\biggl(0,\frac{1}{r}Y_{lm,\theta}(\theta,\phi),\frac{1}{r\sin\theta}Y_{lm,\phi}(\theta,\phi)\biggr)\right.\nonumber\\
\left.+B_3^{lm} \left(0,-\frac{1}{r}\frac{Y_{lm,\phi}(\theta,\phi)}{\sin{\theta}},\frac{1}{r\sin\theta}\sin{\theta}\,Y_{lm,\theta}(\theta,\phi)\right)\right],
\end{eqnarray}
where each vector component represents the \(r,\,\theta,\,\phi\) component in polar coordinates, \(Y_{lm}(\theta, \phi)\) is the \(l,\,m\) component of the spherical harmonics, comma denotes the partial derivative (\(_{,\theta}=\partial_\theta,\,\, _{,\phi}=\partial_\phi\)), and \(B^{lm}_1\), \(B^{lm}_2\), and \(B^{lm}_3\) are functions of \(t\) and \(r\).
 Using these expressions, we rewrite Maxwell equations in terms of \(B_i^{lm}\) and \(E_i^{lm}\,(i=1,\,2,\,3)\) in polar coordinates.
 The time evolution of the magnetic fields, \(B_1^{lm}(t,\,r)\) and \(B_3^{lm}(t,\,r)\), obeys \citep{RufTio1972}
\begin{eqnarray}
\partial_t^{~2}\tilde{B}_1^{lm}(t,r)&=&\partial_r^2\tilde{B}^{lm}_1(t,r)-\frac{l(l+1)}{r^2}\tilde{B}^{lm}_1(t,\,r)-4\pi l(l+1)J_3^{lm}(t,r),\label{eq:B1}\\
\partial_t^{~2}B_3^{lm}(t,r)&=&\partial_r^2B^{lm}_3(t,r)-\frac{l(l+1)}{r^2}B^{lm}_3(t,\,r)-4\pi\left[J_1^{lm}(t,r)-\partial_r\left(J_2^{lm}(t,r)\right)\right],\label{eq:B3}
\end{eqnarray}
where
\begin{equation}
\tilde{B}_1^{lm}(t,r)=r^2 B_1^{lm}(t,r),
\end{equation}
and \(J_i^{lm}\,(i=1, 2, 3)\) are the vector spherical harmonics expansions of the current term.
 Using the condition of $\mathrm{div}\vec{B}=0$, \(B_2^{lm}(t,\,r)\) is related to \(B_1^{lm}(t,\,r)\) as
\begin{equation}
B_2^{lm}(t,\,r)=\frac{1}{l(l+1)}\partial_r\tilde{B}_1^{lm}(t,\,r).
\label{eq:B2}
\end{equation}

Using Amp\'{e}re--Maxwell's law, the evolution equation for the electric field is written as
\begin{eqnarray}
\partial_tE_1^{lm}(t,r)&=&-\frac{l(l+1)}{r^2}B_3^{lm}(t,r)-4\pi J_1^{lm}(t,r),\label{eq:E1}\\
\partial_tE_2^{lm}(t,r)&=&-\partial_r\left(B_3^{lm}(t,r)\right)-4\pi J_2^{lm}(t,r),\label{eq:E2}\\
\partial_tE_3^{lm}(t,r)&=&\frac{1}{l(l+1)}\partial_t^{~2}\tilde{B}_1^{lm}(t,r).\label{eq:E3}
\end{eqnarray}

%dipole current
The current term of a magnetic dipole moment is written as \citep{Dix1970II}
\begin{equation}
J^\mu(x)=-\nabla_\lambda\,\int d\tau \,m^{\lambda\mu}\frac{\delta^{(4)}(x^\nu-x_{\rm{s}}^\nu(\tau))}{\sqrt{-g}},
\label{source}
\end{equation}
where \(m^{\lambda \mu}\) is a magnetization tensor, \(\nabla_\lambda\) is a covariant derivative, \(x_{\rm{s}}^\nu(\tau)\) is the position of the magnetic dipole moment, \(\tau\) is its proper time of it,  \(\delta^{(4)}(x^\mu)\) is a delta function in \(4\) dimension, and \(g\) is the determinant of the metric \(g_{\mu\nu}\).
 We assume that the magnetic dipole vector is parallel to the angular momentum vector of the orbital motion.
 In spherical polar coordinates, the position of the NS, \(x^\mu_{\rm{s}}(\tau)\), is written as 
\begin{equation}
x^\mu_{\rm{s}}(\tau)=\left(t(\tau),\,a,\,\frac{\pi}{2},\,\Omega t(\tau)\right).
\end{equation} 
 Then each component of the magnetization tensor \(m^{\lambda\mu}\) is related to the $z$-component of the magnetic dipole moment \(m^z\) as
\begin{eqnarray}
m_{tr}=-m_{rt}&=&m^zr\Omega \dot{t}_{\rm NS}\\
m_{r\phi}=-m_{\phi r}&=&m^zr\dot{t}_{\rm NS},\\
m_{ij}&=&0\quad\quad\quad({\rm others}).
\end{eqnarray}
where the dot denotes a derivative with respect to the proper time \(\tau\).
Here, we used $m_{\mu\nu}=\dot{x}_s^\rho(\tau)\epsilon_{\rho\mu\nu\sigma}m^\sigma$ where $m^\sigma$ is magnetic dipole moment vector and $\epsilon_{\mu\nu\sigma\rho}$ is the antisymmetric tensor in 4 dimension.
In this context, the source term of Eq.~(\ref{source}) is expanded in vector spherical harmonics as:
\begin{eqnarray}
J_1^{lm}(t,\,r)&=&\frac{m^z}{|\dot{t}|^2}\frac{im}{a^3}Y_{lm}^*\left(\frac{\pi}{2},\Omega t\right)\delta(r-a),\label{eqj1}\\
J_2^{lm}(t, \,r)&=&\frac{m^{z} i m}{l(l+1)} Y_{lm }^{*}\left(\frac{\pi}{2},\Omega t\right)  \partial_{r}\left[\frac{1}{r} \delta(r-a)\right],\label{eqj2}\\
J_3^{lm}(t, \,r)&=&-\frac{m^{z}}{l(l+1)}\,\partial_{\theta} Y_{l m}^{*}\left(\frac{\pi}{2}, \Omega t\right) \,\partial_r\left[\frac{1}{r} \delta(r-a)\right].\label{eqj3}
\end{eqnarray}
where we used $\dot{t}=\left(1-a^2\Omega^2\right)^{-1/2}$.

%FT
We use Fourier transformation to solve the Maxwell equations.
 The Fourier transformation of the time dependent equations of the magnetic field, Eqs.~(\ref{eq:B1}) and (\ref{eq:B3}), is
\begin{equation}
\partial_r^2 B^{lm}(\omega, r)+\left(\omega^2-\frac{l(l+1)}{r^2}\right)B^{lm}(\omega, r)=S^{lm}(\omega, r),
\label{eq:BS}
\end{equation}
where \(B^{lm}(\omega, r)\) is the Fourier transformation of \(\tilde{B}^{lm}_1(t,r)\) or \(B^{lm}_3(t,r)\) and \(S^{lm}(\omega,r)\) is the Fourier transformation of \(4\pi l(l+1)J_3^{lm}(t,r)\) or \(4\pi\left[J_1^{lm}(t,r)-\partial_r\left(J_2^{lm}(t,r)\right)\right]\), respectively.
 We note that for $m=0$, $\omega=0$ in Eq.~(\ref{eq:BS}).
 Equation~(\ref{eq:BS}) is a Strum-Liouville equation and we can solve it by imposing appropriate boundary conditions and constructing Green's functions. 

%solving
To solve Eq.~(\ref{eq:BS}), we impose the boundary condition that the magnetic field is outgoing at infinity and regular at the origin for $m\neq0$ modes, and the magnetic field is regular at infinity and at the origin for $m=0$ modes.
 For $m\neq0$, a homogeneous solution of Eq.~(\ref{eq:BS}) with the outgoing boundary condition at infinity is \(\omega r \,h^{(1)}_l(\omega r)\), and one with the regular boundary condition at the origin is \(\omega r \,j_l(\omega r)\).
 Here, \(j_l(x)\) is the spherical Bessel function of the first kind and \(h_l^{(1)}\) is the spherical Hankel function of the first kind.
 With these homogeneous solutions, we obtain the Green's function of Eq.~(\ref{eq:BS}), \(G_{\omega,l}(r,r')\), for each frequency \(\omega\,(\neq0)\) as,
\begin{eqnarray}
G_{\omega,l}(r,r')=\begin{cases}
-ir'\,h_l^{(1)}(\omega r')\cdot \omega r\,j_l(\omega r) & (r<r')\\
-ir'\,j_l(\omega r')\cdot \omega r\,h^{(1)}_l(\omega r) & (r>r').
\end{cases}
\label{eqgreen}
\end{eqnarray}
For $m=0$, a homogeneous solution with the regular boundary condition at infinity is $r^{-l}$, and one with the regular boundary condition at the origin is $r^{l+1}$.
 Thus, the Green's function for $m=0$ modes is 
 \begin{eqnarray}
G_{\omega,l}(r,r')=\begin{cases}
-\frac{r'}{2l+1}\left(\frac{r}{r'}\right)^{l+1} & (r<r')\\
-\frac{r'}{2l+1}\left(\frac{r'}{r}\right)^l & (r>r').
\end{cases}
\label{eqgreen0}
\end{eqnarray}
We do not consider any effect of the companion on the electromagnetic field.
 If the companion is a BH, we need to impose such boundary condition that the electromagnetic field is ingoing on the event horizon of the BH.
 If the companion is an NS, we need to impose some boundary condition for the magnetic field on the surface of the NS.
 In this paper, we ignore such effects.

 %EMfieldn
With these Green's functions, we derive $B_1^{lm}(\omega,r)$ and $B_3^{lm}(\omega,r)$ in Fourier space as
\begin{eqnarray}
B_1^{lm}(\omega, r)&=&-i\frac{4\pi m^z}{r^2} Y_{lm,\theta}^*\left(\frac{\pi}{2}, 0\right)\delta(\omega-m\Omega)\left\{\theta(r-a)\left[\frac{1}{r'}\partial_{r'}\left(r'\,j_l(\omega r')\right)\right]_{r'=a}\right.\nonumber\\
& &\times\,\omega r\,h_l^{(1)}(\omega r)\left.+\theta(a-r)\left[\frac{1}{r'}\partial_{r'}\left(r'\,h^{(1)}_l(\omega r')\right)\right]_{r'=a}\,\omega r\,j_l(\omega r)\right\},\\
B_3^{lm}(\omega, r)&=&-4\pi m^z m \Omega^2 Y_{lm}^*\left(\frac{\pi}{2}, 0\right)\delta(\omega-m\Omega)\nonumber\\
& &\times\left\{\theta(r-a)\,\left(1-\frac{m^2}{l(l+1)}\right) j_l(\omega a) \,\omega r\,h_l^{(1)}(\omega r) \right.\nonumber\\
& &\left.+\theta(a-r)\,\left(1-\frac{m^2}{l(l+1)}\right) h_l^{(1)}(\omega a)\,\omega r\,j_l(\omega r)+\frac{i}{a\Omega^2 l(l+1)}\delta(r-a)\right\},
\end{eqnarray}
for $m\neq0$ modes. 
 For $m=0$ modes, the solutions are
\begin{eqnarray}
B_1^{l0}(\omega,r)&=&4\pi m^z\frac{1}{ar^2}Y^*_{l0,\theta}\delta(\omega)\left(\frac{\pi}{2},0\right)\left\{\theta(a-r)\frac{l}{2l+1}\left(\frac{r}{a}\right)^{l+1}\right.\nonumber\\
& &\left.-\theta(r-a)\frac{l+1}{2l+1}\left(\frac{a}{r}\right)^l\right\},\\
B_3^{l0}(\omega,r)&=&0.
\end{eqnarray}

%solution fot m neq0
For $m\neq0$ modes, by inverse Fourier transformation of the solution with Eq.~(\ref{eq:B2}), we obtain the electromagnetic field in terms of the vector spherical harmonics expansion as:
\begin{eqnarray}
B_1^{lm}(t,r)&=&-i\frac{4\pi m^z}{r^2} Y_{lm,\theta}^*\left(\frac{\pi}{2}, 0\right)e^{-im\Omega t}\nonumber\\
& &\times\left\{\theta(r-a)\left[\frac{1}{r'}\partial_{r'}\left(r'\,j_l(m\Omega r')\right)\right]_{r'=a} m\Omega r\,h_l^{(1)}(m\Omega r)\right.\nonumber\\
& &\left.+\theta(a-r)\left[\frac{1}{r'}\partial_{r'}\left(r'\,h^{(1)}_l(m\Omega a')\right)\right]_{r'=a}\,m\Omega r\,j_l(m\Omega r)\right\}\label{eq:solb1},\\
B_2^{lm}(t,r)&=&-i\frac{4\pi m^z}{l(l+1)} Y_{lm,\theta}^*\left(\frac{\pi}{2}, 0\right)e^{-im\Omega t}\nonumber\\
& &\times\left\{\theta(r-a)\left[\frac{1}{r'}\partial_{r'}\left(r'\,j_l(m\Omega r')\right)\right]_{r'=a}\frac{d}{dr}\left[m\Omega r\,h_l^{(1)}(m\Omega r)\right]\right.\nonumber\\
& &\left.+\theta(a-r)\left[\frac{1}{r'}\partial_{r'}\left(r'\,h^{(1)}_l(m\Omega r')\right)\right]_{r'=a}\,\frac{d}{dr}\left[m\Omega r\,j_l(m\Omega r)\right]-\frac{i}{r}\delta(r-a)\right\} \label{eq:solb2},\\
B_3^{lm}(t,r)&=&-4\pi m^z m \Omega^2 Y_{lm}^*\left(\frac{\pi}{2}, 0\right)e^{-im\Omega t}\left\{\theta(r-a)\,\left(1-\frac{m^2}{l(l+1)}\right)\right.\nonumber\\
& &\left.\times j_l(m\Omega a) \,m\Omega r\,h_l^{(1)}(m\Omega r) +\theta(a-r)\,\left(1-\frac{m^2}{l(l+1)}\right) h_l^{(1)}(m\Omega a)\,m\Omega r\,j_l(m\Omega r) \right.\nonumber\\
& &\left.+\frac{i}{a\Omega^2 l(l+1)}\delta(r-a)\right\},
\end{eqnarray}
where \(\theta(x)\) is a step function.
 Using this magnetic field and Eqs.~(\ref{eq:E1})--(\ref{eq:E3}), we obtain the electric field as:
\begin{eqnarray}
E_1^{lm}(t,r)&=&\frac{4\pi i m^z l(l+1)}{r^2} \Omega Y_{lm}^*\left(\frac{\pi}{2}, 0\right)e^{-im\Omega t}\nonumber\\
& &\times\left\{\theta(r-a)\,\left(1-\frac{m^2}{l(l+1)}\right) j_l(m\Omega a) \,m\Omega r\,h_l^{(1)}(m\Omega r) \right.\nonumber\\
& &\left.+\theta(a-r)\,\left(1-\frac{m^2}{l(l+1)}\right) h_l^{(1)}(m\Omega a)\,m\Omega r\,j_l(m\Omega r)+\frac{i a^3}{l(l+1)}\delta(r-a)\right\},\label{eq:sole1}\\
E_2^{lm}(t,r)&=&4\pi i m^z \Omega Y_{lm}^*\left(\frac{\pi}{2}, 0\right)e^{-im\Omega t}\nonumber\\
& &\times\left\{\theta(r-a)\,\left(1-\frac{m^2}{l(l+1)}\right) j_l(m\Omega a) \,\frac{d}{dr}\left[m\Omega r\,h_l^{(1)}(m\Omega r)\right] \right.\nonumber\\
& &\left.+\theta(a-r)\,\left(1-\frac{m^2}{l(l+1)}\right) h_l^{(1)}(m\Omega a)\,\frac{d}{dr}\left[m\Omega r\,j_l(m\Omega r)\right]\right\} \label{eq:sole2}\\
E_3^{lm}(t,r)&=&-\frac{4\pi m^z m\Omega}{l(l+1)} Y_{lm,\theta}^*\left(\frac{\pi}{2}, 0\right)e^{-im\Omega t}\left\{\theta(r-a)\left[\frac{1}{r'}\partial_{r'}\left(r'\,j_l(m\Omega r')\right)\right]_{r'=a}\right.\nonumber\\
& &\left.\times\,m\Omega r\,h_l^{(1)}(m\Omega r)+\theta(a-r)\left[\frac{1}{r'}\partial_{r'}\left(r'\,h^{(1)}_l(m\Omega r')\right)\right]_{r'=a}\,m\Omega r\,j_l(m\Omega r)\right\}\label{eq:sole3}.
\end{eqnarray}
For $m=0$ modes, we obtain the magnetic field as:
\begin{eqnarray}
B_1^{l0}(r)&=&4\pi m^z\frac{1}{a}Y^*_{l0,\theta}\left(\frac{\pi}{2},0\right)\left\{\theta(a-r)\frac{l}{2l+1}\left(\frac{r}{a}\right)^{l+1}-\theta(r-a)\frac{l+1}{2l+1}\left(\frac{a}{r}\right)^l\right\},\\
B_2^{l0}(r)&=&4\pi m^z\frac{1}{a}Y^*_{l0,\theta}\left(\frac{\pi}{2},0\right)\left\{\theta(a-r)\frac{1}{2l+1}\left(\frac{r}{a}\right)^l\frac{1}{a}+\theta(r-a)\frac{1}{2l+1}\left(\frac{a}{r}\right)^{l+1}\frac{1}{a}\right.\nonumber\\
& &\left.-\delta(r-a)\right\},\\
B_3^{l0}(r)&=&0\label{eq:solb30}.
\end{eqnarray}
We note that the electric field does not contain the $m=0$ modes because they are static modes and the electric field is induced only by the orbital motion of the magnetic dipole moment.

%%%
%subsection
\subsection{Magnetic field configuration}
%intro
In this subsection, we display the configuration of the magnetic field derived in the previous subsection and illustrate the importance of the higher multipole modes.
 With this configuration, we indicate the location in which the magnetic reconnection could occur in the late inspiral phase if plasma would exist.
 We note that the magnetic reconnection process is associated with the motion of the plasma in the magnetosphere and thus force-free or particle-in-cell simulations are needed for more realistic description.

\begin{figure}[htbp]
  \begin{center}
          \includegraphics[clip, width=15cm]{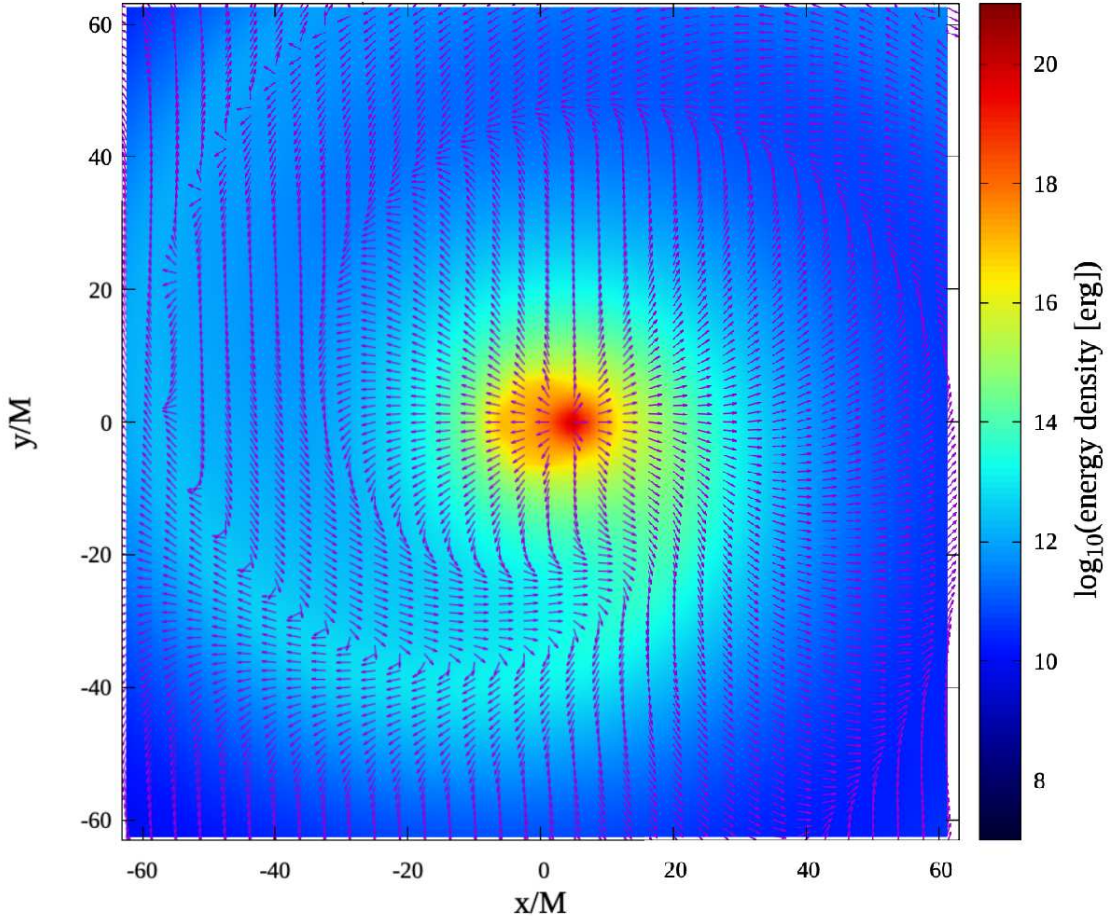}
              \caption{The magnetic-field configuration for \(z=0.5M\) surface. The color shows the strength of the energy density in logarithmic scale, \(\log_{10}{\left(\frac{1}{8\pi}(B^2+E^2)\right)}\), and the arrows show the direction of the magnetic field. The NS is located at \((x,\,y,\,z)=(6M,\,0,\,0)\) and orbiting according to Kepler's law. The center of mass is at the origin, and the magnetic field at the pole of the NS is set to be equal to \(10^{12}\,{\rm{G}}\). We sum higher multipole terms up to $l=85$ modes. We can see the spiral arm configuration, for which the energy density is high and near the spiral arm, the magnetic field direction steeply changes. The convergence of multipole expansion is poor near $x^2+y^2+(0.5M)^2\sim(6M)^2$, so we interpolate the magnetic field there.}
    \label{fig:Bxy}
  \end{center}
\end{figure}

\begin{figure}[htbp]
  \begin{center}
          \includegraphics[clip, width=15cm]{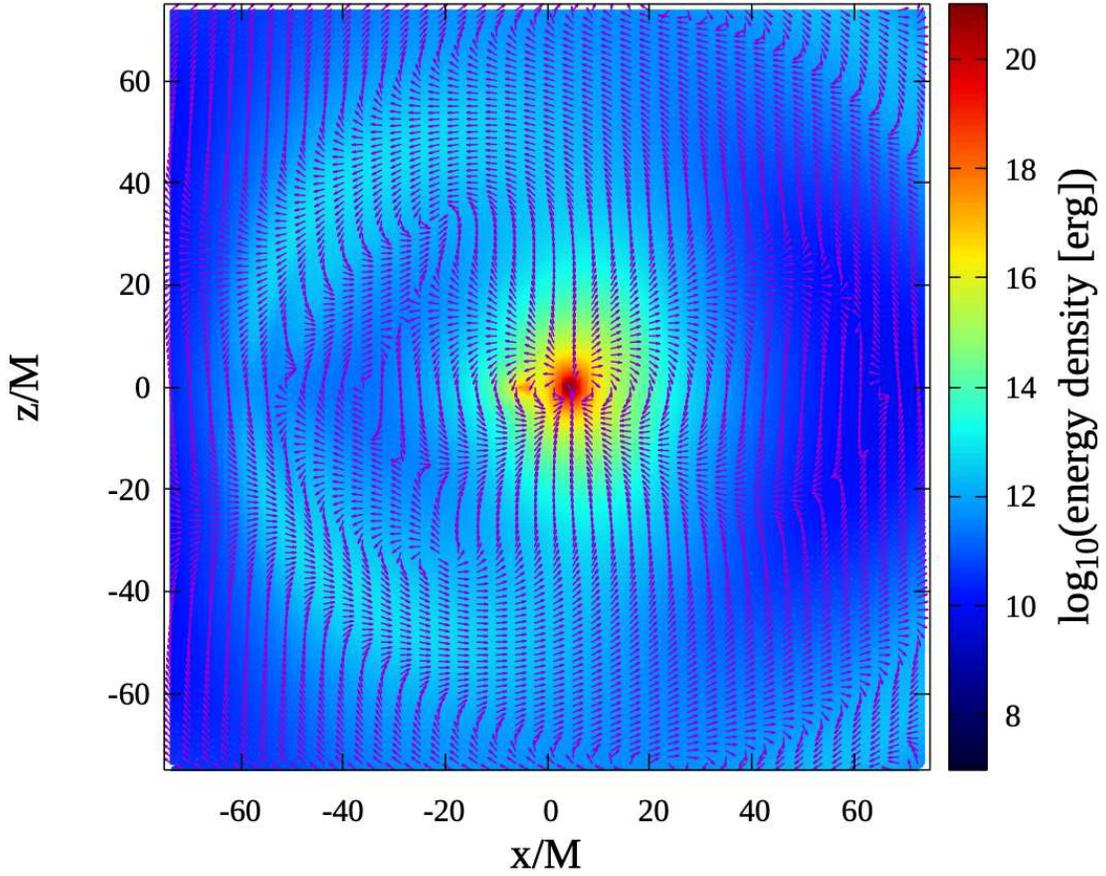}
              \caption{The same as Fig.\ref{fig:Bxy} but for the \(y=0\) surface. Near the NS, the magnetic field is dipolar and near the spiral arm the magnetic field direction steeply changes (displayed in Fig.~\ref{fig:Bxy}, also see Fig.~\ref{fig:costheta}). This region is a candidate for the magnetic reconnection in the presence of plasma.}
             \label{fig:Bxz}
    \end{center}
\end{figure}

\begin{figure}[htbp]
\vspace{-3cm}          
  \begin{center}
          \includegraphics[clip, width=\textwidth]{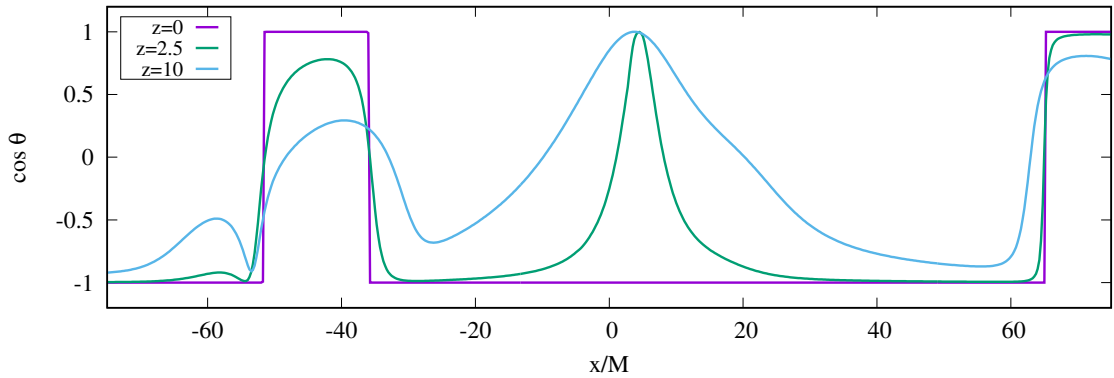}
        \vspace{-3.5cm}          
              \caption{The cosine of the angle between the $z$-axis and the magnetic field in $(y,\,z)=(0,\,0)$ (purple line), $(y,\,z)=(0,\,2.5M)$ (green line), and $(y,\,z)=(0,\,10M)$ (blue line). Here, $\cos\theta=\vec{B}\cdot\vec{e}_z/|\vec{B}|$ where $\vec{e}_z$ is the unit vector in the $z$ direction. Near $x/M\sim-35$, $x/M\sim-50$, and $x/M\sim65$, the magnetic field direction steeply changes.}
             \label{fig:costheta}
    \end{center}
\end{figure}

%field
Figures \ref{fig:Bxy} and \ref{fig:Bxz} show the magnetic field configuration and the energy density of the electromagnetic field for a BH-NS binary just before the merger for \(z=0.5M\) surface and \(y=0\) surface, respectively.
 Here, the NS is located at $(6M,\,0,\,0)$.
 It is found that there is a spiral arm in which the energy density of the electromagnetic field is significantly enhanced.
 As we will see below, this configuration is due to the superposition of several multipole modes.
 In the spiral arm, the poloidal magnetic field often changes its direction (see Fig.~\ref{fig:costheta}).
 In the presence of the plasma, the magnetic reconnection could occur in the region where the magnetic field direction changes steeply.
 In the spiral arm (for example, $(x,\,z)\sim (-35M,0),\,(-50M,0)$ in Fig.~\ref{fig:Bxz}), in particular, the magnetic reconnection could occur. 
 This speculation is confirmed by the force-free simulation \citep{CarShi2020}.

\begin{figure}[htbp]
        \begin{center}
          \includegraphics[clip, width=15cm]{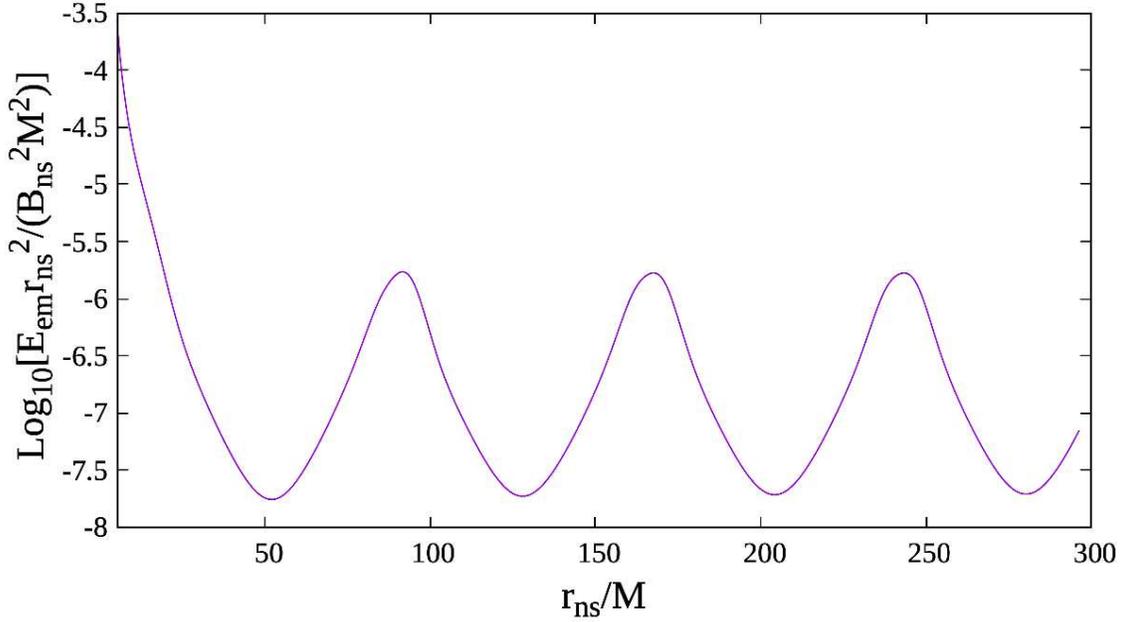}
        \end{center}
    \caption{The energy density times radius squared, \(\frac{1}{8\pi}(E^2+B^2) r_{\rm NS}^2\) , in logarithmic scale at $(r_{\rm NS}/\sqrt{2},\,0,\,r_{\rm NS}/\sqrt{2})$. The left edge of the $r$ axis is at the point where the NS exists. Near the NS the magnetic field is dipolar and the energy density is proportional to \(r^{-6}\). For $r>c/\Omega+a=22M$, the electromagnetic field is composed primarily of electromagnetic waves and the energy density is proportional to \(r^{-2}\) in average. We see  one  peak at each spiral arm and it is due to the summation of the multiple modes.}
    \label{fig:Bline}
\end{figure}

%liner
Figure \ref{fig:Bline} shows the energy density of the electromagnetic field at $(r_{\rm NS}/\sqrt{2},\,0,\,r_{\rm NS}/\sqrt{2})$ where $r_{\rm NS}=\sqrt{(x-6M)^2+y^2+z^2}$  is the distance measured from the NS. 
 The field strength is inversely proportional to $r$ in average for $r\gg R$ and the waveform does not have a sine-wave shape.
 This wave pattern is formed because of the dominant contribution to the field strength from not only the dipole mode but also the higher multipole modes.
 In the case of gravitational waves from binary compact stars, the $l=m=2$ mode is dominant and the waveform has a sine-wave shape.
 However, the wave form of the electromagnetic wave is different from that of the gravitational wave.
 Due to the higher multipole modes of the electromagnetic field, there appear sharp peaks of the energy density and these peaks correspond to the position of  the spiral arm shown in Fig. \ref{fig:Bxy}.
 In the next subsection we show that the contribution of the higher \(l,\,m\) modes enhances Poynting flux.

%%%
%%%subsection
\subsection{Luminosity and special relativistic correction}
\begin{figure}[tbp]
  \begin{center}
          \includegraphics[clip, width=13cm]{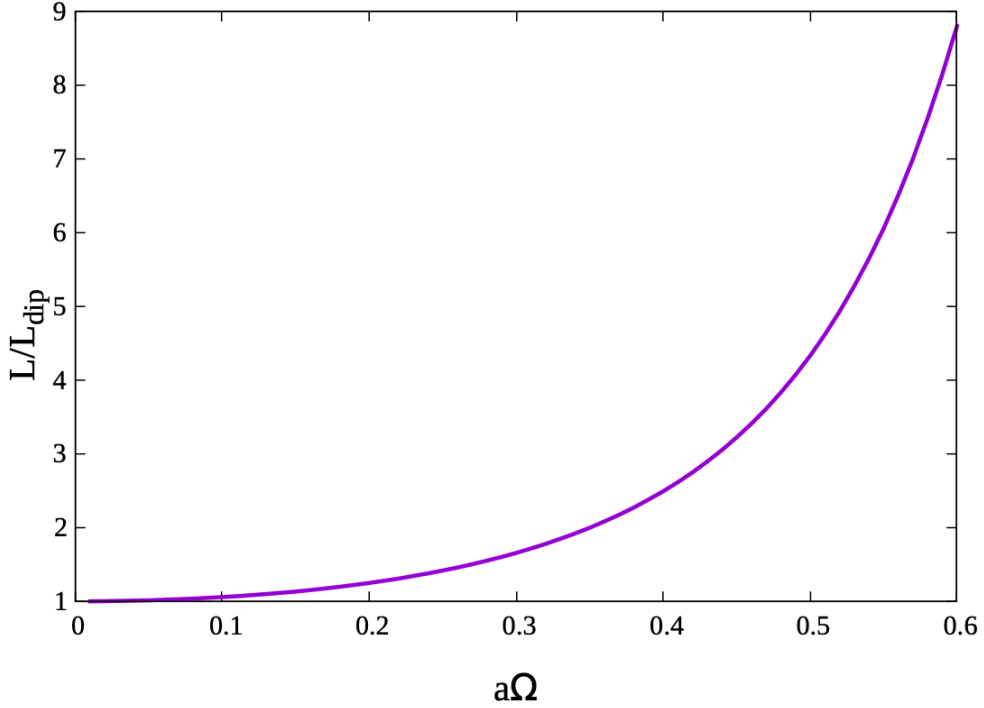}
              \caption{The special relativistic correction for the luminosity, calculated by the analytic formula (\ref{eq:Lana}). At the ISCO of a Schwarzschild BH, the orbital velocity is \(a\Omega=\sqrt{M_{\rm BH}/6M}<0.41\). Therefore, the special relativistic correction would be less than \(2.6\). }
    \label{fig:vL}
  \end{center}
\end{figure}

%luminocity calculation
In terms of  \(B_i^{lm}\) and \(E_i^{lm}\,(i=1,\,2,\,3)\), the Poynting flux at infinity is written as
\begin{eqnarray}
L_{\rm{EM}}&=&\int d\vec{S}\cdot\left(\frac{1}{4\pi}\vec{E}\times\vec{B}\right)\nonumber\\
&=&\int \frac{d\Omega}{4\pi}\sum_{lm\,l'm'}\left[Y_{lm,\theta}Y_{l'm',\theta}\left(E_2^{lm}B_3^{l'm'}-B_2^{lm}E_3^{l'm'}\right)\right.\nonumber\\
& &\left.+\frac{1}{\sin^2{\theta}}Y_{lm,\phi}Y_{l'm',\phi}\left(-E_3^{lm}B_2^{l'm'}+E_2^{lm}B_3^{l'm'}\right)\right.\nonumber\\
& &\left.+\frac{1}{\sin{\theta}}Y_{lm,\theta}Y_{l'm',\phi}\left(E_2^{lm}B_2^{l'm'}-B_2^{lm}E_2^{l'm'}+E_3^{llm}B_3^{l'm'}-B_3^{lm}E_3^{l'm'}\right)\right]\nonumber\\
&=&\frac{1}{4\pi}\sum_{l\,m}(-1)^m\,l(l+1)\left(E_2^{lm}B_3^{l\,-m}-B_2^{lm}E_3^{l\,-m}\right).
\label{eq:L}
\end{eqnarray}
Considering the asymptotic form of the spherical Hankel functions of the first kind at infinity, \(h^{(1)}_l(x)\to \exp{(i(x-\pi(l+1)/2))}/x\), we obtain an analytic formula of the luminosity as
\begin{eqnarray}
L_{\rm{EM}}&=&\frac{4(m^z)^2\Omega^6a^2}{15}\,\frac{15\pi}{a^2\Omega^2}\sum_{lm}\,l(l+1)\left[m^2\left|Y_{lm}\left(\frac{\pi}{2},\,0\right)\right|^2\left(1-\frac{m^2}{l(l+1)}\right)^2\left(j_l\left(m\Omega a\right)\right)^2\right.\nonumber\\
& &\left.+\frac{m^4}{l^2(l+1)^2}\left|Y_{lm,\theta}\left(\frac{\pi}{2},\,0\right)\right|^2\left(\frac{j_l\left(m\Omega a\right)}{m\Omega a}+ j_l'\left(m\Omega a\right)\right)^2\right],
\label{eq:Lana}
\end{eqnarray}
where $j_l'(x)=d j_l(x)/dx$.
 The first term, \(\frac{4(m^z)^2\Omega^6a^2}{15}\), agrees with the luminosity of the dipole radiation derived in Ioka and Taniguchi (2000) \citep{IokTan2000}, and the remaining terms correspond to the special relativistic correction to the radiation formula due to the higher multipole modes of the electromagnetic field. 
 Figure \ref{fig:vL} shows the special relativistic correction of this analytic luminosity.
 We suppose the quasi-circular orbit of the binary terminates at an ISCO around a Schwarzschild BH.
 In the following discussion, we approximately set the radius of the ISCO to be $r\,(=aM/M_{\rm BH})=6M_{\rm BH}$.
 At the ISCO of the BH, the velocity \(a \Omega\) of the NS is $\sqrt{M_{\rm BH}/6M}$, and the special relativistic correction is less than \(2.6\).
 The luminosity is enhanced by the higher multipole modes and this enhancement is unique to electromagnetic waves.
 In the case of gravitational waves from compact binaries, the higher multipole modes reduce the luminosity. 

%expansion
We expand Eq.~(\ref{eq:Lana}) in terms of \(a\Omega\), perform the summation with respect to \(l,\,m\), and obtain \(L_{\rm{EM}}\) in the series of the velocity \(v=a\Omega\) as
\begin{eqnarray}
L_{\rm{EM}}&=&\frac{4(m^z)^2 \Omega^6 a^2}{15}\left[1+\frac{11 v^2}{2}+16 v^4+35 v^6+65 v^8+\frac{217 v^{10}}{2}+168 v^{12}+246 v^{14}\right.\nonumber\\
& &\left.+345 v^{16}+\frac{935 v^{18}}{2}+616 v^{20}+O\left((a\Omega)^{22}\right)\right].
\label{eq:LSR}
\end{eqnarray}

%kiyo of each component 
Each \(l,\,m\) mode contributes substantially to the luminosity.
 To reveal the contribution of each \(l,\,m\) mode, we write the luminosity as
\begin{equation}
L_{\rm{EM}}=\frac{4(m^z)^2 \Omega^6 a^2}{15}\sum_{l,m}\eta_{l,m},
\end{equation}
where \(\eta_{l,m}\) is the Poynting flux of the \(l,\,m\) mode divided by the luminosity at the zeroth order, \(4(m^z)^2 \Omega^6 a^2/15\), and it expresses the contribution of the $l,\,m$ mode in the Poynting flux.
 We show \(\eta_{l,m}\) only for \(m>0\) because \(\eta_{l,m}=\eta_{l,-m}\).
\begin{eqnarray}
\eta_{1,1}&=&\frac{5}{16}-\frac{v^2}{16}+\frac{3v^4}{560}-\frac{v^6}{3780}+\frac{v^8}{116424}-\frac{v^{10}}{5045040}+\frac{v^{12}}{291891600}-\frac{v^{14}}{21709437750}\nonumber\\
& &+\frac{v^{16}}{2016565551000}-\frac{v^{18}}{228678533483400}+\frac{v^{20}}{31079491596153000}+O\left(v^{22}\right),
\end{eqnarray}
\begin{eqnarray}
\eta_{2,1}&=&\frac{3}{16}-\frac{5v^2}{112}+\frac{31 v^4}{7056}-\frac{v^6}{4158}+\frac{115v^8}{13621608}-\frac{47 v^{10}}{227026800}+\frac{41v^{12}}{10916745840}\nonumber\\
& &-\frac{v^{14}}{19037506950}+\frac{193v^{16}}{330313437253800}-\frac{v^{18}}{188806378927320}\nonumber\\
& &+\frac{31v^{20}}{776987289903825000}+O\left(v^{22}\right),\\
\eta_{2,2}&=&\frac{4 v^2}{3}-\frac{16v^4}{21}+\frac{256 v^6}{1323}-\frac{1280 v^8}{43659}+\frac{5120v^{10}}{1702701}-\frac{4096 v^{12}}{18243225}+\frac{131072v^{14}}{10234449225}\nonumber\\
& &-\frac{262144 v^{16}}{453727248975}+\frac{524288 v^{18}}{24773507794035}-\frac{2097152 v^{20}}{3263346617596065}+O\left(v^{22}\right),
\end{eqnarray}
\begin{eqnarray}
\eta_{3,1}&=&\frac{121v^4}{26880}-\frac{121 v^6}{241920}+\frac{11 v^8}{435456}-\frac{11v^{10}}{14152320}+\frac{v^{12}}{60652800}-\frac{v^{14}}{3866616000}\nonumber\\
& &+\frac{v^{16}}{318351384000}-\frac{v^{18}}{32758357413600}+\frac{11v^{20}}{45206533230768000}+O\left(v^{22}\right),\\
\eta_{3,2}&=&\frac{32 v^2}{21}-\frac{64 v^4}{63}+\frac{608 v^6}{2079}-\frac{3968v^8}{81081}+\frac{72704 v^{10}}{13378365}-\frac{210944v^{12}}{487354725}+\frac{9404416 v^{14}}{361129851225}\nonumber\\
& &-\frac{5636096v^{16}}{4587686628525}+\frac{133169152v^{18}}{2848953396314025}-\frac{85458944 v^{20}}{58274046742786875}+O\left(v^{22}\right),\\
\eta_{3,3}&=&\frac{6561v^4}{1792}-\frac{6561 v^6}{1792}+\frac{32805 v^8}{19712}-\frac{59049v^{10}}{128128}+\frac{177147 v^{12}}{2013440}-\frac{531441v^{14}}{42785600}\nonumber\\
& &+\frac{14348907 v^{16}}{10568043200}-\frac{43046721v^{18}}{362483881760}+\frac{129140163 v^{20}}{15158416873600}+O\left(v^{22}\right),
\end{eqnarray}
\begin{eqnarray}
\eta_{4,1}&=&\frac{5v^4}{37632}-\frac{v^6}{59136}+\frac{283 v^8}{295975680}-\frac{43v^{10}}{1331890560}+\frac{311 v^{12}}{420496876800}-\frac{173v^{14}}{13981521153600}\nonumber\\
& &+\frac{233 v^{16}}{1468059721128000}-\frac{191v^{18}}{118178807550804000}+\frac{v^{20}}{74546928696240000}+O\left(v^{22}\right),\\
\eta_{4,2}&=&\frac{512 v^6}{6615}-\frac{2048 v^8}{72765}+\frac{16384 v^{10}}{3468465}-\frac{32768 v^{12}}{66891825}+\frac{524288v^{14}}{14783093325}\nonumber\\
& &-\frac{2097152 v^{16}}{1092306340125}+\frac{16777216v^{18}}{206445898283625}-\frac{8388608 v^{20}}{3021617238514875}+O\left(v^{22}\right),\\
\eta_{4,3}&=&\frac{10935 v^4}{1792}-\frac{19683v^6}{2816}+\frac{50132601 v^8}{14094080}-\frac{7617321v^{10}}{7047040}+\frac{495834453 v^{12}}{2224851200}-\frac{2482360911v^{14}}{73976302400}\nonumber\\
& &+\frac{10029885993v^{16}}{2589170584000}-\frac{73997313399v^{18}}{208428232012000}+\frac{3486784401 v^{20}}{131476064720000}+O\left(v^{22}\right),\\
\eta_{4,4}&=&\frac{8192 v^6}{945}-\frac{131072v^8}{10395}+\frac{4194304 v^{10}}{495495}-\frac{33554432  v^{12}}{9555975}+\frac{2147483648 v^{14}}{2111870475}\nonumber\\
& &-\frac{34359738368v^{16}}{156043762875}+\frac{1099511627776v^{18}}{29492271183375}-\frac{2199023255552 v^{20}}{431659605502125}+O\left(v^{22}\right),
\end{eqnarray}
\begin{eqnarray}
\eta_{5,1}&=&\frac{841 v^8}{670602240}-\frac{841v^{10}}{8717829120}+\frac{841 v^{12}}{242853811200}-\frac{841v^{14}}{10837351324800}\nonumber\\
& &+\frac{841 v^{16}}{686365583904000}-\frac{841v^{18}}{57654709047936000}+\frac{841 v^{20}}{6148088519384448000}+O\left(v^{22}\right),\\
\eta_{5,2}&=&\frac{64 v^6}{10395}-\frac{1024 v^8}{405405}+\frac{7424v^{10}}{15810795}-\frac{19456 v^{12}}{366522975}+\frac{950272v^{14}}{229809905325}\nonumber\\
& &-\frac{16384 v^{16}}{68746552875}+\frac{4784128v^{18}}{449278364910375}-\frac{34340864 v^{20}}{90059890420670625}+O\left(v^{22}\right),\\
\eta_{5,3}&=&\frac{45927 v^8}{112640}-\frac{413343v^{10}}{1464320}+\frac{8680203 v^{12}}{95180800}-\frac{3720087v^{14}}{202259200}\nonumber\\
& &+\frac{100442349 v^{16}}{38429248000}-\frac{43046721v^{18}}{153716992000}+\frac{129140163 v^{20}}{5463940352000}+O\left(v^{22}\right),
\end{eqnarray}
\begin{eqnarray}
\eta_{5,4}&=&\frac{65536 v^6}{3465}-\frac{4194304 v^8}{135135}+\frac{121634816v^{10}}{5270265}-\frac{1275068416 v^{12}}{122174325}+\frac{249108103168v^{14}}{76603301775}\nonumber\\
& &-\frac{17179869184v^{16}}{22915517625}+\frac{20066087206912v^{18}}{149759454970125}-\frac{576144092954624 v^{20}}{30019963473556875}+O\left(v^{22}\right),\\
\eta_{5,5}&=&\frac{244140625 v^8}{12773376}-\frac{6103515625 v^{10}}{166053888}+\frac{30517578125 v^{12}}{925157376}-\frac{762939453125v^{14}}{41285147904}\nonumber\\
& &+\frac{3814697265625v^{16}}{522945206784}-\frac{95367431640625v^{18}}{43927397369856}+\frac{2384185791015625 v^{20}}{4684257919531008}\nonumber\\
& &+O\left(v^{22}\right)
\end{eqnarray}

\begin{figure}[htbp]
  \begin{center}
    \begin{tabular}{c}
      \begin{minipage}{0.5\hsize}
        \begin{center}
          \includegraphics[clip, width=7.5cm]{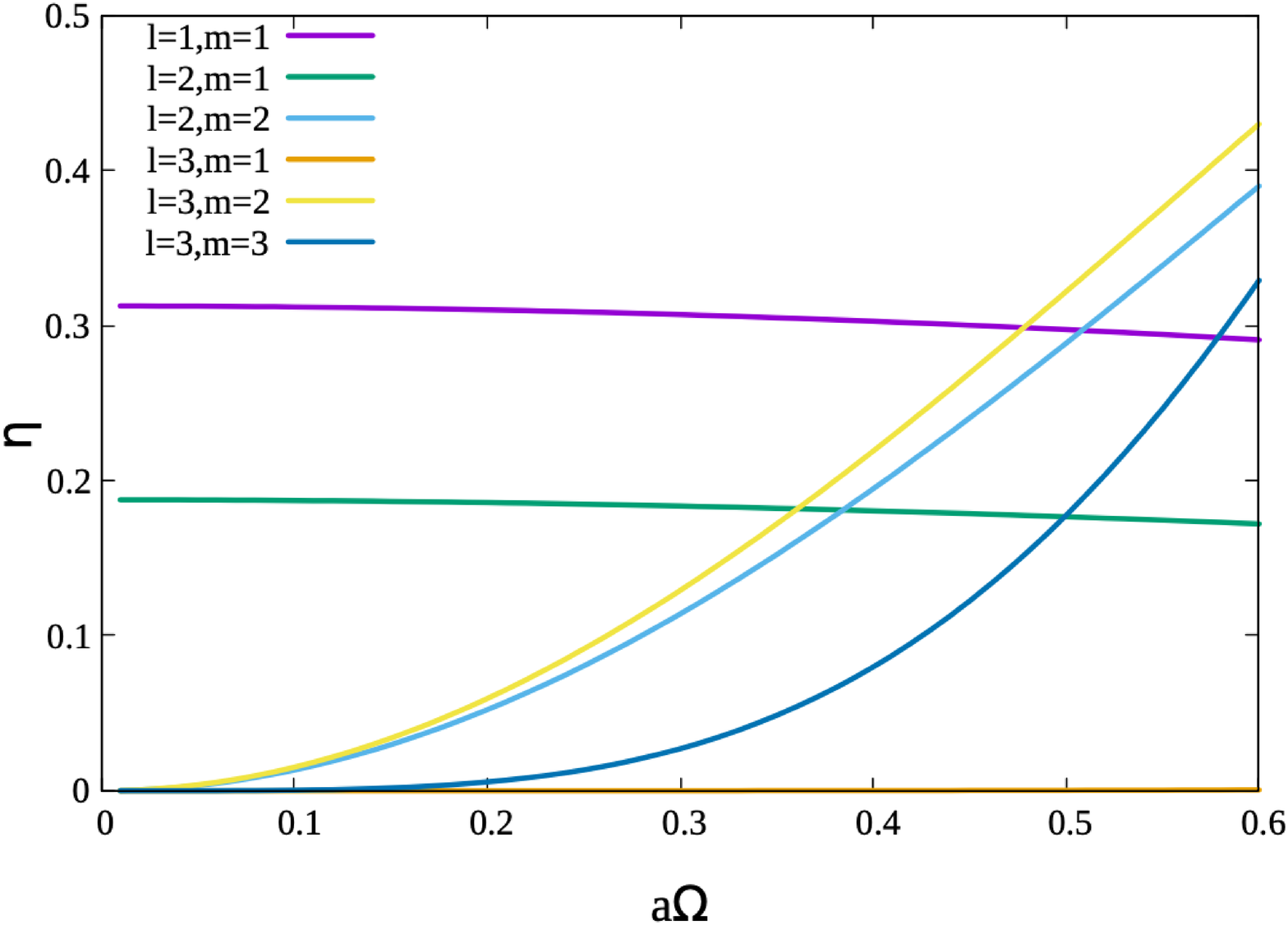}
        \end{center}
      \end{minipage}
      \begin{minipage}{0.5\hsize}
        \begin{center}
          \includegraphics[clip, width=7.5cm]{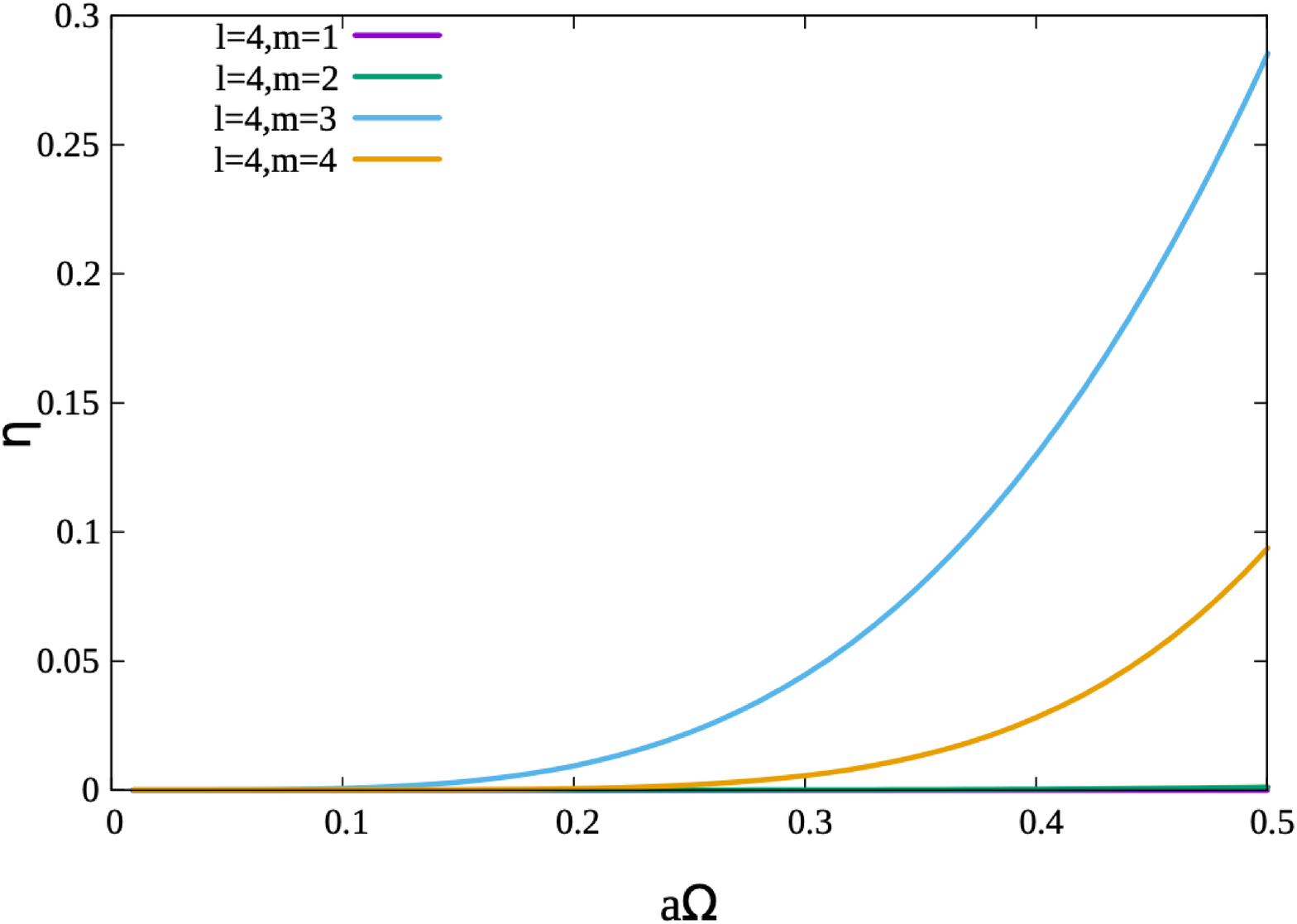}
        \end{center}
      \end{minipage}
    \end{tabular}
    \caption{The velocity dependence of \(\eta_{lm}\). The left panel shows for \(l=1,\,2,\,3\), and the right panel shows for \(l=4\).}
    \label{fig:L1234}
  \end{center}
\end{figure}

\begin{figure}[htbp]
  \begin{center}
    \begin{tabular}{c}
      \begin{minipage}{0.5\hsize}
        \begin{center}
          \includegraphics[clip, width=7.5cm]{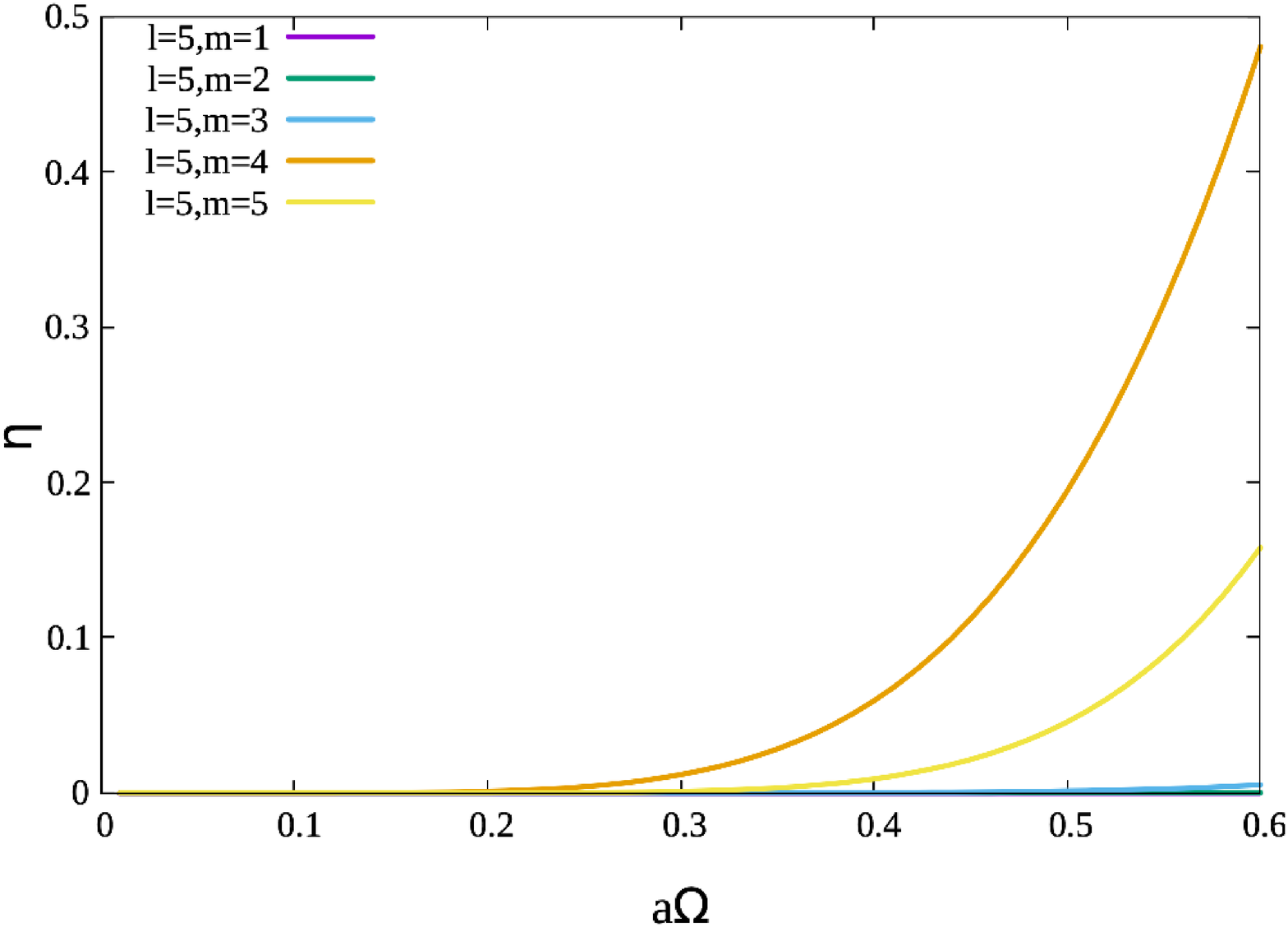}
        \end{center}
      \end{minipage}
      \begin{minipage}{0.5\hsize}
        \begin{center}
          \includegraphics[clip, width=7.5cm]{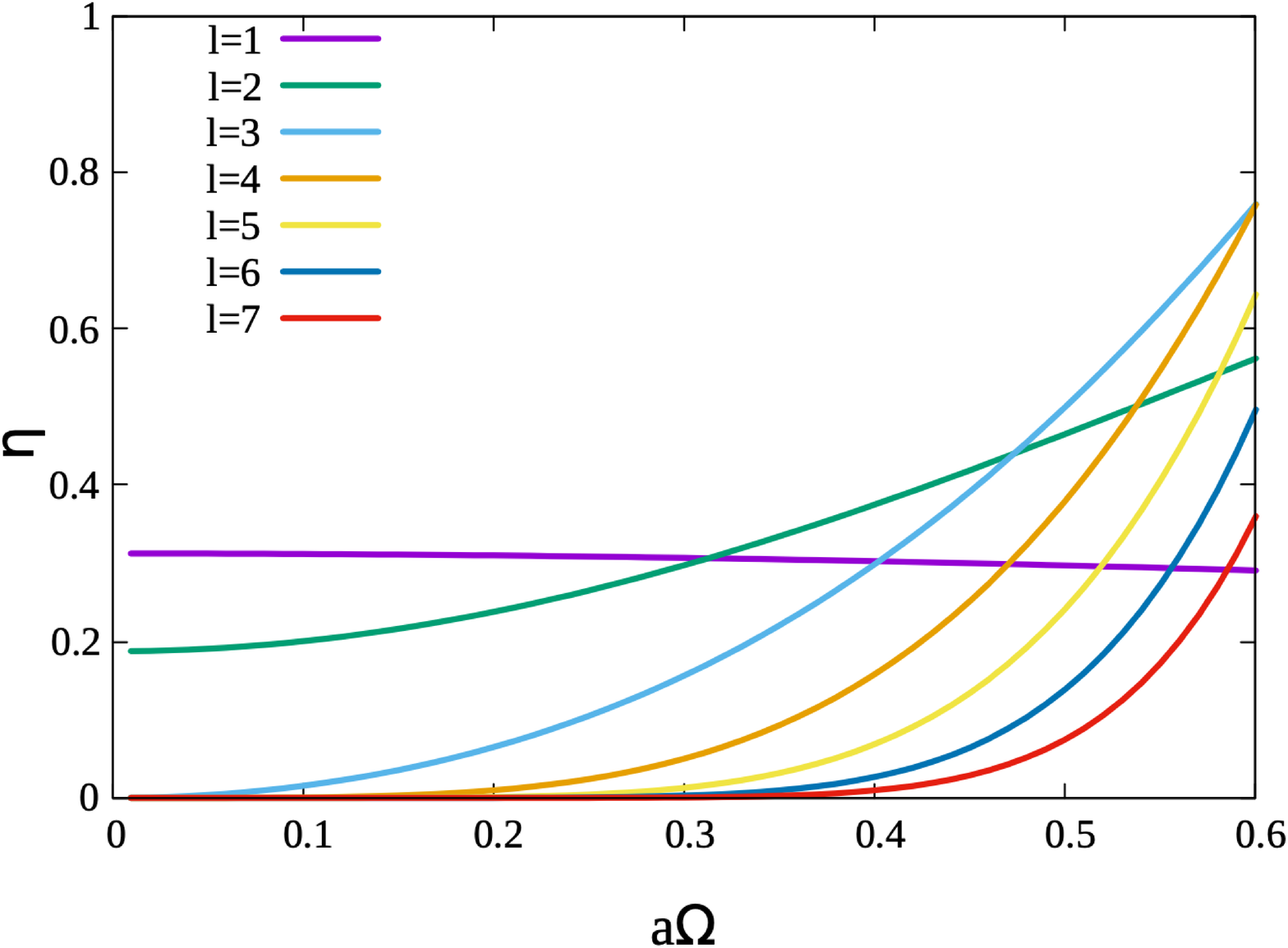}
        \end{center}
      \end{minipage}
    \end{tabular}
    \caption{The velocity dependence of \(\eta_{lm}\) for \(l=5\) modes (left panel). The velocity dependence of \(\sum_{m=1}^l=\eta_{lm}\) for $l=1$--7  (right panel).}
    \label{fig:L5leqm}
  \end{center}
\end{figure}

%Figure meaning
Figures \ref{fig:L1234} and \ref{fig:L5leqm} show the dependence of \(\eta_{lm}\) on the velocity, \(a\Omega\).
 The left panel of Fig.~\ref{fig:L1234} shows \(l=1,2,3\) modes, the right panel of Fig.~\ref{fig:L1234} shows \(l=4\) mode, and the left panel of Fig.~\ref{fig:L5leqm} shows \(l=5\) mode. 
 It is found that for each \(l\), the higher $m$ modes become dominant for higher velocity.
 The right panel of Fig.~\ref{fig:L5leqm} shows the contribution of each $l$ mode (the summation of $m$ modes) for $l=$1--7 modes. 
 This figure indicates that the contribution of \(l >7\) modes would be subdominant in a few milliseconds before the merger. 

%explain 1st order formula
At zeroth order of \(v/c\), the luminosity is 
\begin{eqnarray}
L&=&\frac{4}{15c^5}\left(\frac{B_{\rm{NS}}R_{\rm{NS}}^3}{2}\right)^2\Omega^6a^2,
\label{eq:Lzeroth}
\end{eqnarray}
and, apart from the difference in the numerical coefficient and the dependence on the inclination angle, this is different from the dipole radiation formula for pulsars, $\sim B_{\rm NS}^2R_{\rm NS}^6\Omega^4/c^3$, by an additional factor of $(a\Omega/c)^2$.
 Here, we recover the speed of light, $c$, for clarification of the physical dimension.
 This difference is understood as follows.
 The time dependent part of the magnetic field is of the order of \(B_{\rm{t\h dep}}\sim B_{\rm{dip}}\,a\Omega/c\), which is the difference between the dipole magnetic field and the Lorentz boosted magnetic field.
 The wave field becomes comparable to the dipole field at \(r\sim c/\Omega\).
 At these points, the energy density, $\varepsilon_{\rm{mag}}$,  is 
\begin{equation} 
\varepsilon_{\rm{mag}}\sim\frac{1}{8\pi}\left(B_{\rm{t\h dep}}\left(\frac{R_{\rm{NS}}}{c/\Omega}\right)^3\right)^2.
\end{equation} 
This energy density is radiated at the speed of light.
 Therefore the order of the luminosity is 
\begin{equation}
L\sim \varepsilon_{\rm{mag}}\times 4\pi\left(\frac{c}{\Omega}\right)^2\,c \sim \frac{1}{2c^5} B_{\rm{NS}}^2 R_{\rm{NS}}^6\Omega^6a^2.
\end{equation}

%%%
%%%%subsection
\subsection{Electromagnetic field around BNS}
%the calculation details
Using the solution of Maxwell equations in vacuum shown in Eqs.~(\ref{eq:solb1})--(\ref{eq:solb30}), we also calculate the electromagnetic field around BNSs in which both of the NSs have magnetic field of comparable strength.
 For simplicity, we set that both NSs have the same magnetic field strength at their pole and have the same mass.
 The separation is set to be 50 km.
 We calculate two cases.
 In the first case, both NSs have aligned magnetic dipole moments.
 In the second case, both NSs have anti-aligned magnetic dipole moments.
 We consider the magnetic-field profile, which is determined by the superposition of two different magnetic fields induced by each NS.
 One of the fields is calculated using Eqs.~(\ref{eq:solb1})--(\ref{eq:solb30}), which express the magnetic field induced by one of the NSs.
 In addition, due to the symmetry relative to the $z$ axis, the field contribution from the other NS can be calculated by transforming Eqs.~(\ref{eq:solb1})--(\ref{eq:solb30}) with a rotation relative to the $z$ axis by $\pi$.

\begin{figure}[htbp]
  \begin{center}
    \begin{tabular}{c}
      \begin{minipage}{0.5\hsize}
        \begin{center}
          \includegraphics[clip, width=7.5cm]{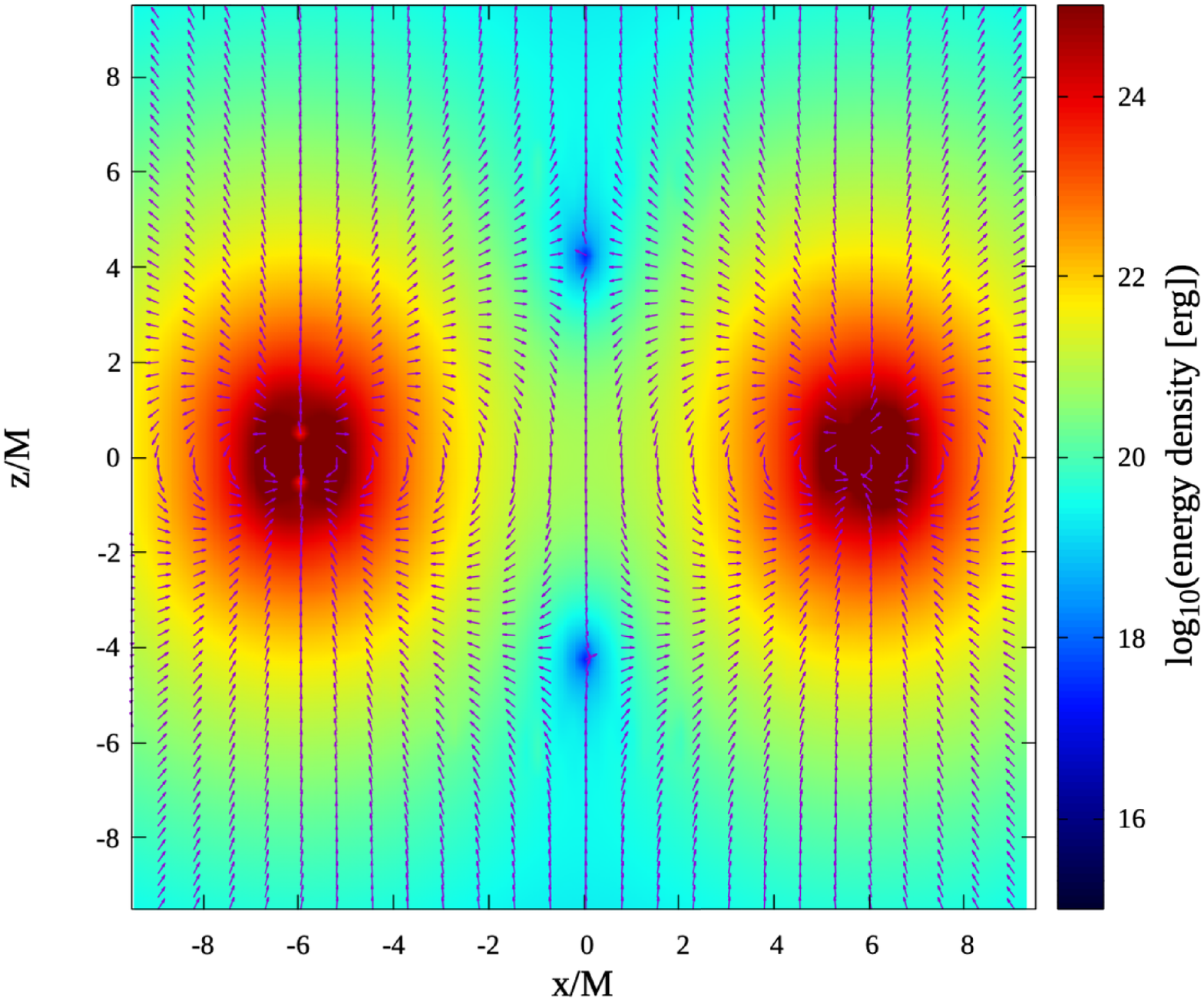}
        \end{center}
      \end{minipage}
      \begin{minipage}{0.5\hsize}
        \begin{center}
          \includegraphics[clip, width=7.5cm]{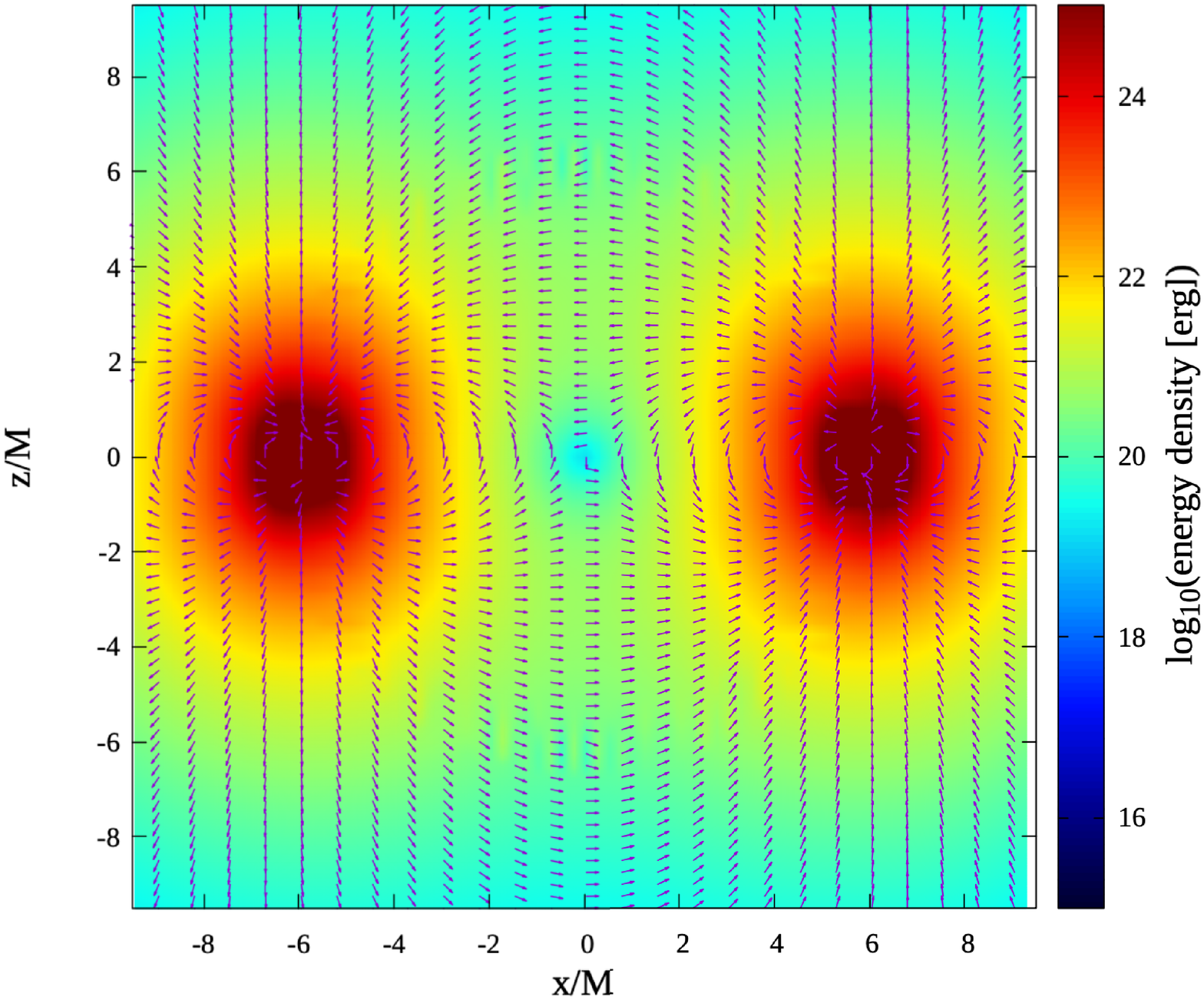}
        \end{center}
      \end{minipage}
    \end{tabular}
    \caption{The magnetic field configuration around a BNS. The left panel shows the magnetic field configuration of aligned magnetic dipole moments. The right panel shows that of anti-aligned ones. The convergence of multipole expansion is poor near $r\sim a$, so we interpolate the magnetic field there.}
    \label{fig:BNSxz}
  \end{center}
\end{figure}

% configuration 
Figure \ref{fig:BNSxz} shows the electromagnetic field around a BNS.
 The left panel of Fig.~\ref{fig:BNSxz} shows the magnetic field configuration of aligned magnetic dipole moments.
 There are two regions around which the magnetic field is weaker.
 For this case, we simply have double spiral arm structure and do not have remarkable difference from the single NS case. 
 The right panel of Fig.~\ref{fig:BNSxz} shows the magnetic field configuration for the anti-aligned case.
 Near the origin, the magnetic field has the opposite direction (X-point) and  magnetic reconnection could occur in the presence of plasma.

The resistive magnetohydrodynamic simulations for BNS in Palenzuela et al. (2013)\citep{PalLeh2013sup} and Palenzuela et al. (2013)\citep{PalLeh2013} show that, for the aligned case, the current sheets first arise at far distances from the stars.
 This current sheet formation might be due to the spiral arm structure of the electromagnetic field.
 They also show that, for the anti-aligned case, the current sheet formation begins between the stars.
 The X-point in the anti-aligned case might trigger the current sheet formation between the stars.

%%%%%%%%%%%%%%
%%%%section
%%%%%%%%%%%%%%
\section{Electromagnetic counterpart of late inspiring neutron-star binaries}
\label{sec:3}
Recently, Carrasco and Shibata (2020) \citep{CarShi2020} showed that an orbiting NS generates a magnetosphere around the NS.
 In this section, we first indicate that the electric field induced by the orbital motion is indeed likely to generate a magnetosphere.
 Then, we consider possible electromagnetic radiation as a precursor and evaluate whether the precursor is observable or not, with the assumption that the total electromagnetic flux is converted into any kind of electromagnetic radiation or particle wind like pulsars.

%%%
%%subsection
\subsection{Magnetosphere formation}
\label{sec:magnetosphere}
%magnetosphere calculation GJ
For an isolated rotating NS (i.e., pulsar), we assume that the NS is a rotating good conductor and the magnetic fields inside and outside the NS are dipole.
 Then an electric field parallel to the dipole magnetic field lines is induced at the surface of the NS, and therefore a magnetosphere is generated \citep{GolJul1969}.
 The charge number density of the magnetosphere (Goldreich-Julian density) is given by Gauss's law under no electric field condition in the comoving frame of plasma.

%magnetosphere formation Binary
In this subsection, we consider the magnetosphere induced by the orbital motion of the NS.
 If the outside of the NS is vacuum, the magnetic dipole moment of the NS accelerated in the orbital motion induces an electric field on the NS surface, and its component parallel to the magnetic field lines is likely to pull out the charged particles from the NS surface.
 The charged particles should be distributed to cancel the induced electric field in the plasma's comoving frame and create a magnetosphere.
 The most interesting feature of this magnetosphere is that it is induced by the orbital motion.
 Thus, the surface charge density may be different from that of an isolated pulsar and the electromagnetic field is not static (e.g., \citep{CarShi2020}).
 We can estimate the charge density of the magnetosphere in a similar way to an isolated pulsar.
 We assume that the plasma in the magnetosphere has a velocity, \(\vec{v}\), in the center-of-mass frame and then the plasma should be distributed to cancel the induced electric field in the plasma's comoving frame.
 Thus, the electric field in the center-of-mass frame is 
\begin{equation}
\vec{E}=-\frac{\vec{v}}{c}\times\vec{B}.
\label{eqEvB}
\end{equation}
Using Gauss's law, we can calculate the charge density in the center-of-mass frame as
\begin{eqnarray}
\rho_c&=&\frac{1}{4\pi}\vec{\nabla}\cdot\vec{E} \nonumber \\
&=&\frac{1}{4\pi c}\left[\left(\vec{\nabla}\times\vec{v}\right)\cdot\vec{B}-\left(\vec{\nabla}\times\vec{B}\right)\cdot\vec{v}\right]\label{eq:rhodef}.
\end{eqnarray}
In the magnetosphere of isolated pulsars with a dipole magnetic field, the ratio of the second term to the first term is of order $\left(v_{\rm rot}/c\right)^2$, where $v_{\rm rot}$ is the rotation velocity of plasmas in the magnetosphere.
This is because $\vec{\nabla}\times\vec{B}=4\pi\rho_c\vec{v}_{\rm rot}/c$ for the stationary pulsar case.
Thus, near the surface, the second term could be neglected and the first term of Eq.~(\ref{eq:rhodef}) mainly contributes to the charge density of the magnetosphere.
We note that in the pulsar case, the second term is proportional to $\rho_c$ and $\rho_c=0$ if the first term vanishes.
 By contrast, for the binary case, \(\vec{\nabla}\times\vec{B}\neq4\pi\rho_c\vec{v}_{\rm rot}/c\) in general due to the orbital motion (i.e., $\partial E/\partial t\neq0$).
 Therefore, even if the NS is irrotating and the motion of the plasma is the same as that of the NS (\(\vec{\nabla}\times\vec{v}=0\)), the second term of Eq.~(\ref{eq:rhodef}) contributes to inducing the charge density of the magnetosphere, leading to \(\rho_c\neq0\).

%density order estimate
We estimate the order of the charge number density of this magnetosphere.
 First, we separate the magnetic field into three parts as  
\begin{equation}
\vec{B}=\vec{B}_{\rm{dip}}+\vec{B}_{\rm{moving}}+\vec{B}_{\rm{rad}},
\end{equation}
where \(\vec{B}_{\rm{dip}},\,\vec{B}_{\rm{moving}},\) and \(\vec{B}_{\rm{rad}}\) denote the pure dipole magnetic field, the modification of the dipole magnetic field due to the orbital motion, and the radiative magnetic field, respectively.
 We note that $\vec{B}_{\rm{moving}}+\vec{B}_{\rm{rad}}$ is equal to $\vec{B}_{\rm t\h dip}$.
 The first and second terms fall off as \(r^{-3}\) and the last one as \(r^{-1}\).
 Rotation of the total magnetic field is nonzero due to the second and third terms,
\begin{equation}
\vec{\nabla}\times\vec{B}=\vec{\nabla}\times\left(\vec{B}_{\rm{moving}}+\vec{B}_{\rm{rad}}\right)\neq\vec{0}.
\end{equation}
Thus, even if the plasma is irrotating i.e., \(\vec{\nabla}\times\vec{v}=0\), Eq.~(\ref{eq:rhodef}) gives \(\rho_c\neq0\), that is, a magnetosphere is induced by the orbital motion.
 By considering the Lorentz transformation of the pure dipole magnetic field \(B_{\rm{dip}}\) with the velocity \(v=a\Omega\), we can estimate the magnetic field strength induced by the orbital motion as \(B_{\rm moving}\sim \left(a\Omega/c\right) B_{\rm{dip}}\).
 With the fact that the magnetic field changes in the length scale of the orbital separation, we estimate $\left|\vec{\nabla}\times\vec{B}\right|\sim B_{\rm moving}/a$, and then, using Eq.~(\ref{eq:rhodef}), the order of the charge density becomes
\begin{equation}
\rho_c\sim-\frac{a \Omega}{4\pi c}\frac{vB_{\rm{dip}}}{ca}=-\frac{B_{\rm{dip}} \Omega}{2\pi c}\frac{a\Omega}{2c}.
\label{eq:app}
\end{equation}
This expression agrees broadly with the result of a force-free simulation in Carrasco and Shibata (2020) \citep{CarShi2020}.

%number density of the magnetosphere
For late inspiral phase, the angular velocity of the orbital motion would become much higher than that of the intrinsic spin, and \(a \Omega\) becomes the order of \(0.1c\).
 Thus, the magnetosphere induced by the orbital motion becomes dominant even in the case of a spinning NS with the spin period 0.1--1 s.
 The particle number density is estimated as
\begin{equation}
n\simeq3.1\times10^{12}\,{\rm{cm^{-3}}}\,\left(\frac{B}{10^{12}\,{\rm{G}}}\right)\left(\frac{M_c}{10M_\odot}\right)\left(\frac{a}{90\,{\rm{km}}}\right)^{-2}.
\label{eq:density}
\end{equation}
This is larger than the Goldreich-Julian density of a typical pulsar with a period \(P=1\,{\rm{s}}\), \(n\sim7\times10^{10}\,{\rm cm^{-3}}\,B_{12}P_0^{-1}\) (here, \(B=10^{12}\,{\rm G}\,B_{12}\) and $P=1\,{\rm s}\,P_0$).

%%%
%%subsection
\subsection{Possible precursor radiation and its observability}
%over view
An accelerated magnetic dipole moment radiates electromagnetic waves as Poynting flux.
 In this subsection, we evaluate the luminosity and the observability of precursors to BNS and BH-NS binary mergers.
 Some kinds of radiations from pulsars are comparable to the total Poynting flux that is estimated by the magnetic dipole emission (for review, \citep{Har2007,Pet2016,CerBel2017}).
 In considering the observability, we assume that the radiation efficiency is approximately equal to that of pulsars suggested by observations \citep{ManHob2005,fermiGRP2010,RayAbd2012,fermiGRP2013,Car2014,OlaKas2014}.

\begin{figure}[tbp]
  \begin{center}
          \includegraphics[clip, width=15cm]{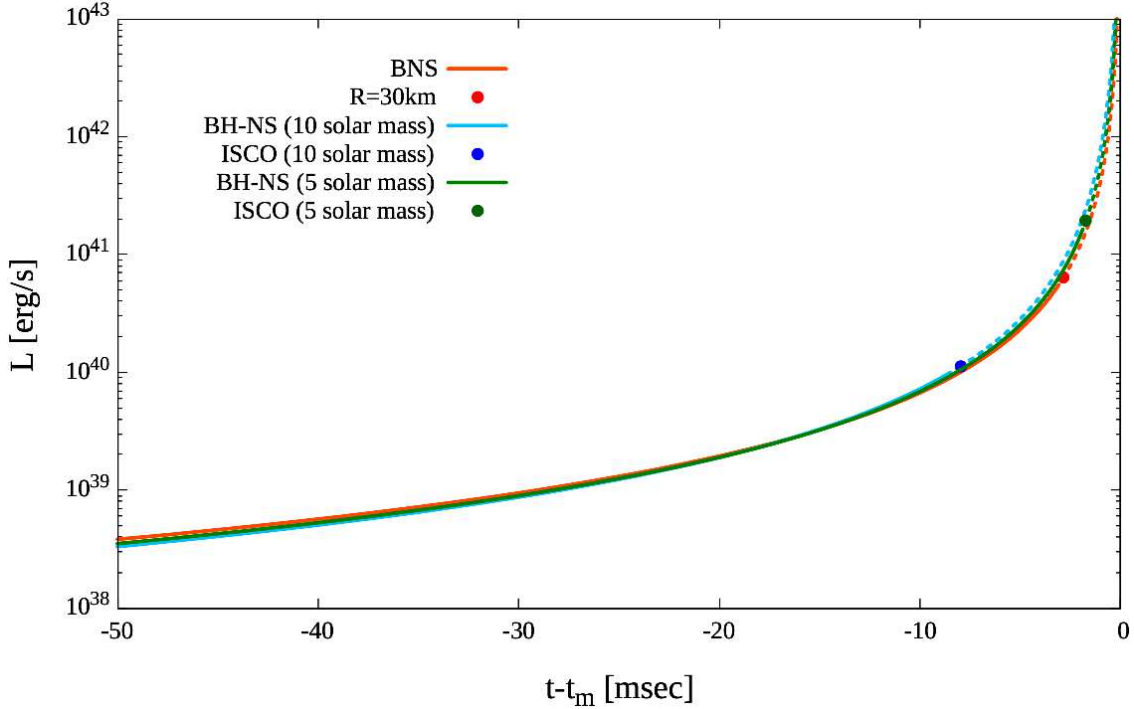}
              \caption{Light curve of the total luminosity. Here, \(t_m\) denotes the merger time. We fixe the magnetic field at the NS pole as \(10^{12}\,{\rm{G}}\), mass of NS as \(1.4M_\odot\), and radius of NS as \(12\,{\rm{km}}\).The BH mass is \(10M_\odot\) and \(5M_\odot\). We assume that the BNS merges when  the separation reaches \(30\,{\rm{km}}\) (red dot). We also assume that the BH-NS binary merges when the separation reaches \(90\,{\rm{km}}\) for the \(10M_\odot\) BH-NS binary case (blue dot), and \(45\,{\rm km}\) for the \(5M_\odot\) BH-NS binary case (green dot).}
    \label{fig:lc}
  \end{center}
\end{figure}

%luminosity from binary NS
First, we consider the luminosity of a BNS just before its merger.
 We set the mass of each NS as \(1.4M_\odot\) and the separation as \(30\,{\rm{km}}\).
 Then, the orbital velocity of the NS is \(0.19c\), and the special relativistic correction is  \(f_r\sim1.2\).
 Using Eq.~(\ref{eq:LSR}), the luminosity is (see Fig.\ref{fig:lc})
\begin{eqnarray}
L_{\rm BNS}&\simeq&5.8\times10^{40}\,{\rm{erg\,s^{-1}}}\,\left(\frac{f_r}{1.2}\right) \left(\frac{B_{\rm{NS}}}{10^{12}\,{\rm{G}}}\right)^2\left(\frac{R_{\rm{NS}}}{12\,{\rm{km}}}\right)^6\left(\frac{M_{\rm{NS}}}{1.4M_\odot}\right)^3\left(\frac{R}{30\,{\rm{km}}}\right)^{-7}.
\end{eqnarray}

%luminosity of BHNS
Next we consider the luminosity of a BH-NS binary in which the NS is at ISCO of the Schwarzschild BH.
 We set the mass of the BH as \(10M_\odot\) and the separation as \(90\,{\rm{km}}\) (ISCO of the Schwarzschild BH).
 The velocity of the NS is \(0.38c\), and the special relativistic correction is  \(f_r\sim2.3\).
 Then the luminosity is (see Fig.\ref{fig:lc}) 
\begin{eqnarray}
L_{\rm{NS\h BH}}\simeq1.0\times10^{40}\,{\rm{erg\,s^{-1}}}& &\,\left(\frac{f_r}{2.3}\right) \left(\frac{B_{\rm{NS}}}{10^{12}\,{\rm{G}}}\right)^2\left(\frac{R_{\rm{NS}}}{12\,{\rm{km}}}\right)^6\nonumber\\
& &\times\left(\frac{M_{\rm{BH}}}{10M_\odot}\right)^2\left(\frac{M_{\rm BH}+M_{\rm NS}}{11.4M_\odot}\right)\left(\frac{R}{90\,{\rm{km}}}\right)^{-7}.
\end{eqnarray}
We note that the orbital radius of ISCO of a BH depends on its spin.
 Thus, if we assume that the spin of the BH is parallel to the orbital angular momentum, the ISCO can be smaller than \(6M_{\rm{BH}}\) down to $M_{\rm BH}$ and $L_{NS\h BH}$ reaches $\sim10^{42}\,{\rm erg\,s^{-1}}$ at 1 ms before the merger.

%set up of estimation
In both cases,  the luminosity just before the merger is typically \(\sim5\times 10^{40}\,{\rm{erg\,s^{-1}}}\).
 Hereafter we use this value as a typical value of the maximum luminosity.
 Also, since the event rate is approximately one BNS merger per year per \(100\,{\rm Mpc}\) cubic volume \citep{LIGO2020GW190425}, we set the luminosity distance as \(100\,{\rm{Mpc}}\), and evaluate the observability of the electromagnetic counterparts.

%%%
%%subsection
\subsubsection{Radio wave emission}
\begin{figure}[tbp]
  \begin{center}
          \includegraphics[clip, width=13cm]{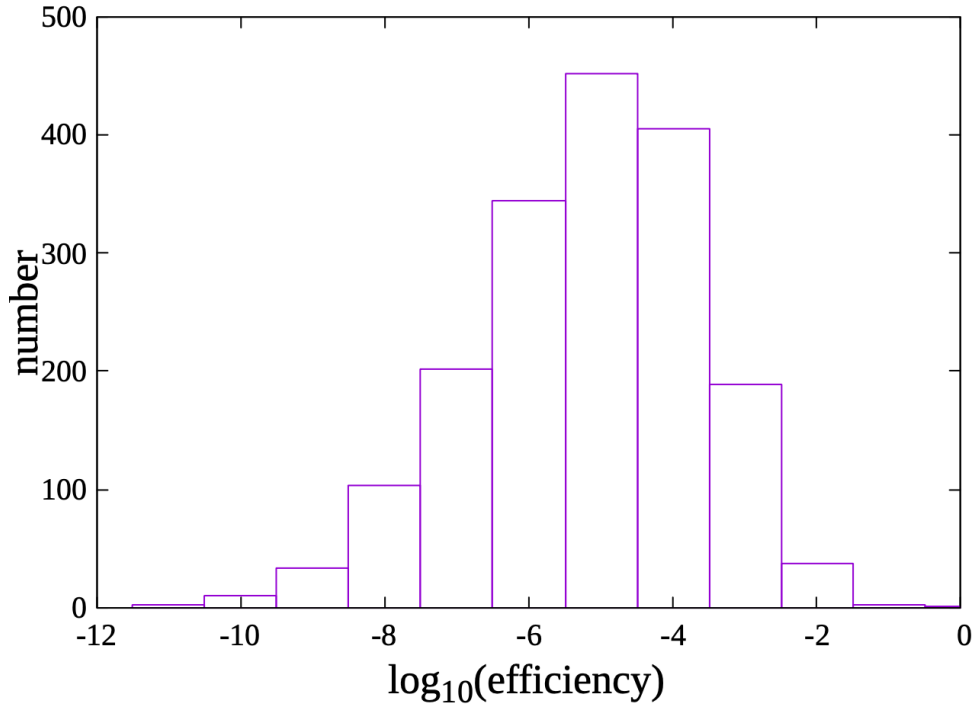}
              \caption{The radio emission efficiency for pulsars from Manchester et al. (2005) \citep{ManHob2005}. The efficiency is defined as the ratio of radio luminosity to spin-down luminosity. We use the radio luminosity at 1.4\,GHz in the ATNF pulsar catalogue and assume that the radio luminosity is equal to $\nu L_\nu$.}
    \label{fig:fraction}
  \end{center}
\end{figure}

%radio wave emission
We first explore the observability of the radio wave emission.
 According to the catalogue of Manchester et al. (2005) \citep{ManHob2005}, the observed spinning-down pulsars emit radio waves and its luminosity is $\epsilon_r\sim10^{-6}$--$10^{-4}$ of its spin-down luminosity (Fig.~\ref{fig:fraction}).
 We note that this catalogue contains all types of pulsars except for accretion-powered systems.
 For example, the catalogue includes gamma-ray pulsars which are quiet in radio \citep{fermiGRP2010,RayAbd2012,fermiGRP2013,Car2014}.
 Some fraction of Poynting flux is radiated as strong coherent emission such as giant pulses of Crab pulsar, and might also power Fast Radio Bursts (FRBs) \citep{LorBai2007,Kat2018,Tot2013}.
 If a radio burst is radiated after the merger of a NS binary, the dynamical ejecta around the merger remnant shields the radio wave \citep{YamTot2017}.
 However, in the precursor case, only the plasma in the magnetosphere is relevant for shielding the radio wave.

%radio wave
If we assume that a fraction \(\epsilon_r\) of the total luminosity goes into radio emission, the spectral flux density at \(\nu=1.4\,{\rm{GHz}}\) is 
\begin{eqnarray}
F_\nu&\simeq&0.5\,{\rm{mJy}}\,\left(\frac{L}{5\times10^{40}\,{\rm{erg\,s^{-1}}}}\right)\left(\frac{\epsilon_r}{10^{-4}}\right)\left(\frac{D}{100\,{\rm{Mpc}}}\right)^{-2}\left(\frac{\nu}{1.4\,{\rm{GHz}}}\right)^{-1}.
\label{eq:Fradio}
\end{eqnarray}
Here, there is uncertainty in \(\epsilon_r\) and \(\nu\).
 With Square Kilometer Array (SKA1) whose sensitivity is \(\sim1\,{\rm mJy}\) for $0.35$--$1.05\,{\rm GHz}$ (SKA1-Mid band 1) and 0.95--1.76 GHz (SKA1-MID band 2) \citep{SKA2016,SKA2019}, we can observe the radio precursor at \(100\,{\rm Mpc}\) for $\epsilon_r > 2 \times10^{-4}$.
 In estimating the sensitivity, we assume that the pulse duration is \(10\,{\rm ms}\) and the band width is \(100\,{\rm MHz}\).
 We note that the field-of-view of SKA will be $\sim200\,{\rm deg^2}$ and approximately $5\%$ of the all sky is covered.
 Therefore, the detection of the radio precursor by chance is unlikely and the prediction of the sky position is important to detect it.
 However, if the BH in a BH-NS binary is rapidly spinning, the prospect for the detection is enhanced because a close orbit is possible and the luminosity can be $\sim10^{42}\,{\rm erg\,s^{-1}}$ at 1 ms before the merger. 
 
 %possibility of FRB
If the value of the efficiency can be as high as \(\epsilon_r\sim0.1\), the flux density is approximately \(F_\nu\sim 0.5\,{\rm{Jy}}\) and this flux density is comparable to typical FRBs.
 The radio wave emission from the magnetosphere induced by the orbital motion can be a candidate for the unknown sources of FRBs if the efficiency for the radio emission is extremely high or the BH is highly spinning.
 We note that the BNS merger event rate is smaller than the FRB event rate, and thus, only a small fraction of the FRBs can be explained by NS mergers \citep{KasIok2013,Rav2019}.

%plasma frequency
The effect of plasma can modify the propagation of radio waves in the magnetosphere induced by the orbital motion.
 We consider the plasma frequency.
 Electromagnetic waves with a frequency below the plasma frequency \(\nu_p\) cannot propagate through the plasma \citep{RybickiLightman1979}.
 Using Eq.~(\ref{eq:density}), the plasma frequency at $r=c/\Omega$ is
\begin{eqnarray}
\nu_p  &\simeq&200\,{\rm{MHz}}\left(\frac{B}{10^{12}\,{\rm{G}}}\right)^{1/2}\left(\frac{M_c}{10M_\odot}\right)^{5/4} \left(\frac{a}{90\,{\rm{km}}}\right)^{-13/4}\\
 &\simeq& 170\,{\rm{MHz}}\left(\frac{B}{10^{12}\,{\rm{G}}}\right)^{1/2}\left(\frac{M_c}{10M_\odot}\right)^{7/16}\left(\frac{M_{\rm{tot}}}{11.4M_\odot}\right)^{-13/16}\left(\frac{t}{10\,{\rm{ms}}}\right)^{-13/16},
\end{eqnarray}
where we set the radius of the NS as \(R_{\rm{NS}}=12\,{\rm{km}}\) and the multiplicity of the magnetosphere as 1.
 Ordinary-mode radio waves, with frequency below \(\nu_p\), cannot propagate through the magnetosphere.
 Because the plasma frequency is proportional to square root of the plasma number density, the ordinary-mode radio wave with the frequency of order 1\,GHz is not able to propagate through the plasma when the multiplicity is larger than $\sim25$.
 Under strong magnetic field, the X-mode of radio waves with frequency $\nu<\nu_p$ can propagate in the plasma.
 This effect can make it easier for low-frequency radio waves to propagate in the magnetosphere.

%free free absorption
Free-free absorption also disturbs the propagation of radio waves.
 For radio waves with the frequency $\nu$ and the size of the source \(l\sim100\,{\rm{km}}\) (this is comparable to $c/\Omega$), we obtain the optical depth for radio waves as \citep{RybickiLightman1979},
\begin{equation}
\tau\simeq6\times10^{-6}\left(\frac{\gamma}{10^2}\right)^{-3/2}\left(\frac{n_e}{10^{12}\,{\rm{cm^{-3}}}}\right)^2\left(\frac{\nu}{1\,{\rm{GHz}}}\right)^{-2}\left(\frac{l}{100\,{\rm{km}}}\right),
\label{eq:tau}
\end{equation}
where $\gamma$ is the Lorentz factor of the plasma particle in the magnetosphere and the multiplicity of the magnetosphere is assumed to be 1.
 The optical depth is smaller than $1$ and thus free-free absorption would not shield the ratio wave.
 We note that if we assume \(\gamma\sim100\), the radio wave with $\nu=1\,{\rm GHz}$ is in the small-angle uncertainty principle range and the velocity averaged Gaunt factor is approximately \(\bar{g}_{\rm{ff}}\sim17\) \citep{NovTho1973}.
 In the situation that the multiplicity is larger than of order $10^3$, the optical depth becomes larger than 1 and the radio wave would be blocked.
 If the motion of plasma is restricted by a strong magnetic field, it is known that the optical depth of the free-free absorption becomes smaller than Eq.(\ref{eq:tau}) \citep{KumLu2017}.
 This effect can make it easier for radio waves to propagate in the magnetosphere.

%emission and polarization
In radio pulsars, the temporal variation of linear polarization position angle of the pulses have information about the projected magnetic field orientation in the emission region.
 With this variation, we can constrain some geometrical parameters of the emission regions such as inclination and radial distance measured from the center of the pulsar \citep{RadCoo1969, DykRud2004}.
 In the binary case, if these kinds of temporal variation are observed, we might be able to constrain the geometrical features of the emission region, such as radial distance, opening angle of emitting cone, and the inclination of magnetic dipole moment with respect to the orbital angular momentum.
 Our analytic solution and the gravitational wave signal can  help to constrain these parameters.

%%%
%%subsection
\subsubsection{Gamma-ray \& X-ray emission}
%gamma and X
Next we consider the observability of X-ray or gamma-ray precursors.
 From pulsar magnetospheres, gamma-rays or X-rays can be emitted.
 Gamma-ray pulsars emit gamma-rays with the fraction from \(10^{-3}\) to \(1\) of the spin down luminosity \citep{fermiGRP2010,RayAbd2012,fermiGRP2013,Car2014}.
% Also magnetars emit X-rays and its luminosity can be higher than the spin-down luminosity since magnetars use their magnetic energy as the emission source \citep{OlaKas2014}.
 Like X-ray emission from ordinary pulsars, binaries just before the merger may emit X-rays and gamma-rays.

%gamma-ray X-ray
If we assume that all photons have the same energy \(\epsilon\), the photon number flux is
\begin{equation}
F_{\rm ph}=2.6\times10^{-7}\,{\rm cm^{-2}s^{-1}}\left(\frac{\epsilon_{X,\gamma}}{1}\right)\left(\frac{L}{5\times10^{40}\,{\rm erg/s}}\right)\left(\frac{\epsilon}{100\,{\rm keV}}\right)^{-1}\left(\frac{D}{100\,{\rm Mpc}}\right)^{-2},
\label{eq:Fph}
\end{equation}
where \(\epsilon_{X,\gamma}\) is the efficiency of X-rays or gamma-rays.
 Therefore, in hard X-ray (\(\sim100\,{\rm keV}\)) , {\rm Swift} Burst Alert Telescope (BAT), whose sensitivity is \(\sim10^{-8}\,{\rm erg\,cm^{-2}s^{-1}}\) at 15--150 keV \citep{swiftBAT2005}, is able to observe the precursor if the strength of the magnetic field \(B\) is stronger than \(\sim10^{14}\,{\rm G}\) and the efficiency \(\epsilon_{\gamma}\) is \(\sim1\).
 Also,  {\it Fermi} Gamma-ray Burst Monitor (GBM), whose trigger sensitivity is \(\sim0.7\,{\rm cm^{-2}s^{-1}}\) at 50--300 keV \citep{fermiGBM2009}, is able to observe the X-ray precursor if the strength of the magnetic field \(B\) is stronger than \(\sim10^{15}\,{\rm G}\) and the efficiency \(\epsilon_{X}\) is \(\sim1\)  or the BH in a BH-NS binary is so rapidly spinning that the orbital radius with $R\sim M_{\rm BH}$ is possible.
 For the cases of the typical magnetic field of NSs, \(B\sim10^{12}\,{\rm{G}}\), and the moderately spinning BH in BH-NS binary, it is difficult to observe the precursor in hard X-rays and gamma-rays.
 Thus, the observation of the precursor in gamma-rays and X-rays is not very likely although this possibility cannot be completely excluded.

%%%
%%subsection
\subsubsection{Particle wind}
%particle wind
The particle-in-cell simulation in Cerutti et al. (2015) \citep{CerPhi2015} shows that the magnetic reconnection at current sheets could accelerate charged particles up to $\gamma\sim\sigma_{\rm LC}$ in a pulsar magnetosphere.
 Here, $\gamma$ is the Lorentz factor of the accelerated particles and \(\sigma_{\rm LC}\) is the sigma parameter of the pulsar at $r\sim c/\Omega$.
 Like the case in this simulation, the charged particles could be accelerated via magnetic reconnection in the magnetosphere induced by the orbital motion and particle wind could be ejected .
 The luminosity of the wind has the same parameter dependence as the Poynting flux, as seen below.

%particle radiation
Here, we estimate the order of the luminosity of the particle wind.
 The accelerated particles are expected to be ejected mainly from the tail part of the spiral arm which is located at $r=c/\Omega$.
 The orbital motion would eject particles from the tail part of the magnetosphere and the ejection would make some vacuum gap in the magnetosphere.
 From Eq.~(\ref{eq:app}), the charge density of the plasma at $r=c/\Omega$ is 
\begin{equation}
\rho_c\simeq\frac{(a \Omega B_{\rm{dip}}) \Omega}{4\pi c^2}\left(\frac{R_{\rm{NS}}}{c/\Omega}\right)^3.
\end{equation}
The total radiated charge per second, \(I\), is \(\sim\rho_c\times4\pi (c/\Omega)^2\,c\).
 We assume that these particles are accelerated by electromotive force \(\Delta\Phi\), and then, the total luminosity is written as 
\begin{equation}
L_{\rm{wind}}\sim\rho_c\,4\pi \left(\frac{c}{\Omega}\right)^2c\,\Delta\Phi.
\end{equation}
Because the system is dynamical, it is not easy to estimate the electromotive force between the magnetic field lines.
 Thus, to know the order of the maximum electromotive force which accelerates the charged particles in the magnetosphere, we assume
\begin{equation}
\Delta\Phi\sim E_{\rm{LC}}\frac{c}{\Omega}\sim \frac{a\Omega}{c}B_{\rm{LC}}\frac{c}{\Omega}\sim a\frac{B_{\rm{dip}}R_{\rm{NS}}^3}{(c/\Omega)^3},
\label{eq:EMforce}
\end{equation}
where $E_{\rm LC}$ denotes the electric field strength at $r=c/\Omega$.
 Then we get the total luminosity as 
\begin{eqnarray}
L_{\rm{wind}}&\sim&6\times10^{40}\,{\rm{erg\,s^{-1}}}\, \left(\frac{B_{\rm{NS}}}{10^{12}\,{\rm{G}}}\right)^2\left(\frac{R_{\rm{NS}}}{12\,{\rm{km}}}\right)^6\left(\frac{M_{\rm c}}{10M_\odot}\right)^2\left(\frac{M_c+M_{\rm NS}}{11.4M_\odot}\right)\left(\frac{R}{90\,{\rm{km}}\,}\right)^{-7}.
\label{eq:Lwind}
\end{eqnarray}
Here $R$ is the separation of the binary.
 This luminosity is comparable to the total Poynting flux in Eq.(\ref{eq:LSR}) and has the same parameter dependence as \(L_{\rm{dip}}\).
 Since the energy of the particle wind increases as the separation of the binary shrinks (see Eq.~(\ref{eq:EMforce})), the particles ejected just before the merger could have higher energy.

%%%%%%%%%%%%%%%%%%%
%%section
%%%%%%%%%%%%%%%%%%%
\section{SUMMARY \& DISCUSSION}
\label{sec:4}
%the configuration of EM field and the effect of higher lm mode
In this paper, we analytically solve the electromagnetic field around binary compact stars containing at least one NS and  study the properties of the electromagnetic field.
 Assuming vacuum, we solve Maxwell equations with the source term of a magnetic dipole moment, using the vector spherical harmonics expansion and Green's function for Strurm-Liouville problem.
 We show that the electromagnetic field is characterized by a spiral arm with the high energy density (see Fig.~\ref{fig:Bxy}).
 This configuration is due to the enhancement of the electromagnetic field not only by the dipole modes but also by the the higher multipole modes.
 The higher multipole modes also enhance the total luminosity at infinity by a factor of 2--4 just before the merger of the binary (see Eq.~(\ref{eq:LSR}) and Fig.~\ref{fig:vL}).
 Also, we show that the direction of the magnetic fields steeply changes at the both edges of the spiral arm (see Figures \ref{fig:Bxy} and \ref{fig:Bxz}).
 In such regions, magnetic reconnection could occur in the presence of plasma, as this possibility is verified in Carrasco and Shibata (2020) \citep{CarShi2020}.

%magnetosphere formation
In addition, we discuss the magnetosphere formation in NS binaries in the late inspiral phase.
 We indicate that a magnetosphere is induced by the orbital motion of magnetized NSs, and estimate the particle number density of the magnetosphere to be approximately \(10^{12}\,{\rm{cm^{-3}}}\) for an NS with the magnetic field at its pole \(10^{12}\,{\rm{G}}\) (see Eq.(\ref{eq:density})).
 This density is higher than the typical particle number density of the magnetosphere induced by the intrinsic spin of the ordinary NSs (\(\sim10^{11}\,{\rm{cm^{-3}}}\) for a pulsar whose magnetic field is \(10^{12}\,{\rm{G}}\) and period is \(1\,{\rm s}\)).
 For ordinary pulsars, the Goldreich-Julian density is induced because the rotation of the velocity of the plasma is not zero (\(\vec{\nabla}\times\vec{v}\neq0\)).
 By contrast, the plasma density of the magnetosphere in NS binaries is induced as a result of the deformation of the magnetic field by the orbital motion, i.e., the acceleration of magnetized NS (\(\vec{\nabla} \times\vec{B}\neq0\)).

%observability 
Using the radiation formula and the number density of the magnetosphere, we estimate the luminosity of the precursor under the assumption that the radiation efficiency is the same as that of ordinal pulsars.
 We consider three possibilities as precursors, particle wind, radio pulse, and gamma-ray \& X-ray emission.
 We evaluate the observability of these radiations, and suggest that gamma-ray \& X-ray emission is observable by the telescopes currently in operation only if the magnetic field of the NS is as strong as that of magnetars or the BH in a BH-NS binary is extremely rapidly spinning.
 Radio waves can be observable if the radiation efficiency \(\epsilon_r\) is extremely high as \(\sim10^{-1}\) for ordinary BNSs or the BH in a BH-NS binary is rapidly spinning, and it might be a candidate for some of FRBs.
 Even if \(\epsilon_r\) is as small as \(2\times10^{-4}\), radio waves can be a candidate for the observation with SKA-MID in the future.

%boundary condition (briefly)
In the original argument in Goldreich and Julian (1969) \citep{GolJul1969}, they first consider a rotating NS in vacuum and show that the induced surface electric charge on the NS creates electric field \(E_\parallel\) parallel to the magnetic field.
 This electric field, \(E_\parallel\), accelerates charged particles from the surface of the NS in vacuum, and the charged particles construct a magnetosphere of the pulsar.
 As this argument, we have to consider whether \(E_\parallel\) exists or not with the surface charge of the NS in vacuum induced by the orbital motion for the discussion of the formation of the magnetosphere.
 After the formation of the magnetosphere, the magnetic field in the NS is likely to be dipolar field and do not contain the other components, $\vec{B}_{\rm moving}$ and $\vec{B}_{\rm rad}$.
 The difference between the magnetic field in the NS and the outer region might be kept by the surface currents on the NS.
 However, for simplicity, we ignore the difference of the magnetic field in the NS and the outer region.
 These currents on the surface of the NS might change the magnetic field around the binary, and the modification might be the power series of \(\sim R_{\rm NS}/R_{\rm LC}\) because this difference is due to the surface condition on the surface of the NS.

%future work, FF, PIC
To establish a more precise picture of the electromagnetic field and the magnetosphere induced by the orbital motion, we need to perform force-free or particle-in-cell simulations.
 The key ingredient of the magnetosphere formation is the orbital motion whose speed is a few\(\times10\)\% of the speed of light in compact orbits.
 Therefore, in a simple head-on collision PIC calculation \citep{CriCer2019}, the configuration of the magnetosphere is not the same as our results.
 Recently, a force-free simulation for a binary is done in Carrasco and Shibata (2020) \citep{CarShi2020}.
 The configuration of the poloidal magnetic field is different from our results because the plasma is present (their Fig.~3).
 In this magnetosphere, Alfv\'{e}n waves are generated and they interact non-linearly, resulting in the formation of the current sheets and dissipation.
 Their result shows the presence of a spiral current sheet and its spiral configuration is the same as the spiral arm in our vacuum solution (their Fig.~2 and our Fig.~\ref{fig:Bxy}).
 According to their Fig.~6 that shows the luminosity, the presence of the plasma enhances the total luminosity by a factor of \(2.5\) from the vacuum luminosity when the velocity of the NS is \(0.4c\).
 Therefore, our vacuum solution can capture the magnetosphere around the orbiting NS at least qualitatively and is able to estimate the luminosity within a factor of $\sim2$ even just before the merger. 
 
 \section*{ACKNOWLEDGMENTS} 
We thank Federico Carrasco for fruitful discussion. We also thank Hamid Hamidani, Wataru Ishizaki, Kazuya Takahashi, Takahiro Tanaka, and Bing Zhang for several helpful discussions. This work is in part supported by JSPS/MEXT grant (No. 16H02183, JP20H00158) (MS), nos. 20H01901, 20H01904, 20H00158, 18H01215, 18H01213, 17H06357, 17H06362, 17H06131 (KI) and nos. 20J13806 (TW).

\end{document}